\documentclass[12pt]{article}
\usepackage{epsfig,amsmath,amssymb,verbatim}
\usepackage[utf8]{inputenc}
\usepackage[colorlinks=true,citecolor=blue,,linktocpage=true,linkcolor=blue,urlcolor=black]{hyperref}
\usepackage{mathtools,slashed}
\usepackage{array}
\usepackage{xcolor}
\usepackage{caption}
\usepackage{subcaption}

\usepackage[left=2.5cm,bottom=3cm,right=2.5cm,top=3cm]{geometry}

\usepackage{overpic,youngtab}
\usepackage[latin,english]{babel}
\usepackage{amsmath}
\usepackage{amssymb}
\usepackage{epstopdf}
\usepackage{graphics,psfrag,rotating}
\usepackage{mathabx}
\usepackage{graphicx}
\usepackage{dcolumn}
\usepackage{float}
\usepackage{pdflscape}
\usepackage{array}
\usepackage{booktabs}
\usepackage{amscd}
\usepackage{mathtools}
\usepackage{fancybox}
\usepackage{tensor}
\usepackage{fix-cm}
\usepackage[colorlinks=true,citecolor=blue,,linktocpage=true,linkcolor=blue,urlcolor=black]{hyperref}
\usepackage{cleveref}
\usepackage{tikz}
\usetikzlibrary{calc}
\usetikzlibrary{intersections}
\usetikzlibrary{positioning}
\usetikzlibrary{arrows}
\usetikzlibrary{shapes.misc}

\usetikzlibrary{patterns,snakes}
\tikzstyle{spring}=[line width=0.8,black,snake=coil,segment amplitude=4.25,segment length=4.75,line cap=round]
\usetikzlibrary{decorations.pathmorphing}
\usetikzlibrary{decorations.markings}
\usetikzlibrary{arrows.meta}
\usetikzlibrary{matrix}
\usetikzlibrary{positioning}
\usetikzlibrary{arrows}
\usepackage{titlesec}
\usepackage{abstract}
\usepackage{cancel}
\usepackage{mathtools}
\usepackage{pdflscape}
\usepackage{graphicx}
\definecolor{mypink}{HTML}{FDA4BA}
\usepackage{pdflscape}
\usepackage{subcaption}

\def\be{\begin{equation}}
\def\ee{\end{equation}}
\def\bea{\begin{eqnarray}}
\def\eea{\end{eqnarray}}
\def\pd{\partial}
\def\a{\alpha}
\def\b{\beta}
\def\g{\gamma}
\def\d{\delta}

\def\m{\mu}
\def\n{\nu}
\def\t{\tau}

\def\l{\lambda}

\def\r{\rho}
\def\G{\Gamma}
\def\fg{\mathfrak{g}}

\def\s{\sigma}

\def\bi{\begin{itemize}}

	\def\ei{\end{itemize}}

\def\xp{x^\prime}

\definecolor{ogreen}{rgb}{0,0.7,0}

\DeclareGraphicsExtensions{.pdf,.png,.jpg,.gif,.jpeg}
\def\be{\begin{equation}}
\def\ee{\end{equation}}
\def\pd{\partial}
\def\a{\alpha}
\def\b{\beta}
\def\g{\gamma}
\def\d{\delta}

\def\m{\mu}
\def\n{\nu}
\def\t{\tau}

\def\l{\lambda}

\def\r{\rho}

\def\s{\sigma}

\def\t{\tau}
\def\bi{\begin{itemize}}
	\def\ei{\end{itemize}}

\def\bpm{\begin{pmatrix}}
\def\epm{\end{pmatrix}}

\def\xp{x^{\prime}}

\usepackage{authblk}
\usepackage{setspace}

\usepackage{caption}
\usepackage{graphicx}
\usepackage{tikz}
\usetikzlibrary{patterns,snakes}
\usetikzlibrary{shapes.misc}
\usetikzlibrary{shapes.geometric}
\usetikzlibrary{decorations.markings,decorations.pathmorphing}
\usetikzlibrary{calc}
\tikzstyle{spring}=[line width=0.8,black,snake=coil,segment amplitude=4.25,segment length=4.75,line cap=round]

\usetikzlibrary{decorations.pathmorphing}
\usetikzlibrary{decorations.markings}
\usetikzlibrary{arrows.meta}
\usetikzlibrary{matrix}
\usetikzlibrary{positioning}
\usetikzlibrary{arrows}

\usetikzlibrary{shapes.misc}

\tikzset{cross/.style={cross out, draw=black, minimum size=2*(#1-\pgflinewidth), inner sep=0pt, outer sep=0pt},
	cross/.default={1pt}}

\begin{document}
	
	\vspace*{-1cm}
\phantom{hep-ph/***}
{\flushleft
	{}
	\hfill{{ IFT-UAM/CSIC-25-15}}}
\vskip 1.5cm
\begin{center}
	{\LARGE\bfseries  One loop analysis of the  cubic action for gravity. }\\[3mm]
	\vskip .3cm
	
\end{center}

\vskip 0.5  cm
\begin{center}
	{\large Enrique \'Alvarez$^{\dagger}$, Jes\'us Anero$^\dagger$, Carmelo P. Martin$^{\dagger\dagger}$ .}
	\\
	\vskip .7cm
	{
		$\dagger$Departamento de F\'isica Te\'orica and Instituto de F\'{\i}sica Te\'orica,
		IFT-UAM/CSIC,\\
		Universidad Aut\'onoma de Madrid, Cantoblanco, 28049, Madrid, Spain\\
        $\dagger\dagger$Universidad Complutense de Madrid (UCM), Departamento de F\'isica Te\'orica and \\
        IPARCOS, Facultad de Ciencias F\'isicas, 28040 Madrid, Spain
		\vskip .1cm

		\vskip .5cm
		
		\begin{minipage}[l]{.9\textwidth}
			\begin{center}
				\textit{E-mail:}
				\tt{enrique.alvarez@uam.es},
				\tt{jesusanero@gmail.com},
                \tt{carmelop@fis.ucm.es}
				
			\end{center}
		\end{minipage}
	}
\end{center}
\thispagestyle{empty}

\begin{abstract}
	\noindent
We analyze some aspects  of the  cubic action for gravity recently proposed by Cheung and Remmen, which  is a particular instance of a first order (Palatini) action. In this approach both the spacetime metric and the connection are treated as independent fields. We discuss its  BRST invariance and compute explicitly the one-loop contribution of  quantum fluctuations around flat space, checking that the corresponding Slavnov-Taylor identities are fulfilled. Finally, our results on a first order action are compared with the existing ones corresponding to a second order action.		
	
\end{abstract}

\newpage
\tableofcontents
\thispagestyle{empty}
\flushbottom

\newpage

\section{From Palatini to Cheung-Remmen.}
One year after the first formulation of General Relativity both Hilbert and Einstein \cite{Einstein}\cite{Hilbert} put forward  a variational principle from which Einstein's equations could be deduced. It is instructive to discuss the subtle differences between them both. What we shall call Hilbert's  action (often called Einstein-Hilbert's) is given by
\be
S_{\text{\tiny{H}}}\left[g_{\m\n}\right]\equiv  -{1\over 2 \kappa^2}\int d^4 x \sqrt{|g|}\,R\left[g_{\m\n}\right]
\ee
(with $\kappa^2\equiv 8 \pi G$) where it is understood that the connection from which the scalar curvature is computed is the Levi-Civita one whose components are  Christoffel's symbols associated to the metric tensor.
\par
Einstein's principle is given by
\be
S_{\text{\tiny{E}}}= -{1\over 2\kappa^2}\int d^4 x\, \sqrt{|g|}g^{\m\n}\left(\Gamma^\l_{\t\m}\Gamma^\t_{\l\n}-\Gamma^\l_{\t\l}\Gamma^\t_{\m\n}\right)
\ee
and it is related to Hilbert's one through
\bea\label{EH} \mathcal{L}_{\text{\tiny{H}}}-\mathcal{L}_{\text{\tiny{E}}}&=\partial_\l\left(\sqrt{|g|} g^{\m\n}\Gamma^\l_{\m\n}\right)-\partial_\n\left(\sqrt{|g|} g^{\m\n}\Gamma^\l_{\l\m}\right)\eea
This means that  Hilbert and Einstein  Lagrangians differ by a total derivative; so that they yield the same equations of motion.  We have suppressed the common factor $ -{1\over 2\kappa^2}$ in our definitions for simplicity.\\

This  Einstein Lagrangian, although not manifestly diffeomorphism invariant,  involves  first derivatives of the metric tensor only, yielding a nice explanation of the (at first sight surprising) fact that Einstein's equations only entail derivatives of the metric tensor up to second order.
\par
In 1919, four years after Einsteins's paper on General Relativity, Attilio Palatini \cite{Palatini} proposed a first order variational principle
\be\label{P}
S_{\text{\tiny{P}}}\left[g_{\m\n},\Gamma^\l_{\a\b}\right]=-{1\over 2 \kappa^2}\int d^n x \sqrt{|g|}g^{\m\n}~R_{\m\n}[\Gamma]
\ee
In this first order Lagrangian  both the metric and the connection (assumed torsionless) are independent fields, so that we cannot integrate by parts covariant derivatives.

We can then write the total variation of Palatini's action
\bea
&&\d S_{\text{\tiny{P}}}=-\frac{1}{2 \kappa^2}\int d^n x \sqrt{|g|}~\bigg\{\left(R_{\m\n }-{1\over 2}R g_{\m\n}\right) \d g^{\m\n}+\nonumber\\
&&+\pd_\r \d \G^\r_{\m\n}-\pd_\n \d \G^\l_{\m\l}+\G^\t_{\s\t} \d \G^\s_{\m\n}+\G^\s_{\m\n}\d \G^\l_{\s\l}-\Gamma^\s_{\r\m}\d \G^\r_{\s\n}-\G^\r_{\n\s} \d \G^\s_{\m\r}\bigg\}
\eea
the corresponding equation of motion

\bea
\bigg\{-\pd_\l \left(\sqrt{|g|} g^{\a\b}\right)+\d_\l^\b \pd_\t\left(\sqrt{|g|} g^{\a\t}\right)+\G^\t_{\l\t} g^{\a\b}+\d^\b_\l g^{\m\n} \G_{\m\n}^\a+ 2 g^{\a\t} \G^\b_{\l\t}\bigg\}\d \G^\l_{\a\b}=0
\eea
implies  that the connection is the Levi-Civita \footnote{
It is important to note that this equivalence does not hold anymore for Lagrangians of higher order in the curvature \cite{AA}; then first order and second order are not equivalent. In general first order is much more general than second order.} one
\be
\G^\l_{\a\b}=\left\{{}^{~\l}_{\a\b}\right\}
\ee

In \cite{Buchbinder},\cite{Alvarez:2017cvx}, it is  claimed (through a background field computation) that the one-loop counterterm in first order (Palatini)  formalism match exactly with the standard result \cite{tHooft} in second order, provided that the gauge fixing term is identical in both cases.\\

\subsection*{The Cheung-Remmen action.}
Start with the Palatini action \eqref{P} and define new convenient variables  $\fg^{\a\b}$ and  $A^\m_{\a\b}$ out of the metric and connection through
\bea\label{def}
&&\fg^{\a\b}\equiv \sqrt{|g|} g^{\a\b}\nonumber\\
&&\frak{g}_{\a\b}=\frac{1}{\sqrt{|g|}}g_{\a\b}\nonumber\\
&&A^\m_{\a\b}\equiv \Gamma^\m_{\a\b}-{1\over 2}\left(\d^\m_\a\Gamma^\l_{\b\l}+\d^\m_\b \Gamma^\l_{\a\l}\right)
\eea
 Palatini's Lagrangian now reads
\bea
&&\mathcal{L}_{\text{\tiny{P}}}\left[\fg^{\m\n},A^\l_{\a\b}\right]=\frak{g}^{\a\b}\left(\partial_\l A^\l_{\a\b}+ \frac{1}{n-1}A^\l_{\l\a}A^\t_{\t\b}-A^\l_{\t\a}A^\t_{\l\b}\right)
\eea
integrating by parts, we obtain  Cheung-Remmen's \cite{Cheung} Lagrangian, which is cubic in the fields
\be\label{CR}
\mathcal{L}_{\text{\tiny{CR}}}[\fg^{\m\n},A^\l_{\a\b}]=-A^\m_{\a\b}\pd_\m\fg^{\a\b}+\fg^{\a\b}\left(\frac{1}{n-1}A^\l_{\l\a}A^\t_{\t\b}-A^\l_{\t\a}A^\t_{\l\b}\right)
\ee
this Lagrangian then shares the virtues of all first order Lagrangians in it being a polynomial in the fundamental fields, $A$ and $\fg$.

\par

This action has recently been instrumental in deriving Schwarzschild's geometry through a clever resummation of Feynman diagrams \cite{Mougiakakos}.
\par
 In fact, in order to study perturbation theory around flat space, it is convenient to make the field redefinition
\bea
&&\frak{g}^{\a\b}=\eta^{\a\b}-h^{\a\b}\nonumber\\
&&A^\l_{\m\n}= B^\l_{\m\n}+\frac{1}{2}\left(\partial_\m h^\l_\n+\partial_\n h^\l_\m-\partial^\l h_{\m\n}+\frac{1}{n-2}\eta_{\m\n}\partial^\l h\right)\eea
Adding   the gauge fixing term
\be \mathcal{L}_{\text{\tiny{GF}}}=-\frac{1}{2}\eta^{\a\b}\partial^\r h_{\r\a}\partial^\s h_{\s\b}\ee
the lagrangian reduces to
 \be \mathcal{L}_{\text{\tiny{EH+GF}}}=\mathcal{L}_{\text{\tiny{hh}}}+\mathcal{L}_{\text{\tiny{BB}}}+\mathcal{L}_{\text{\tiny{hhh}}}+\mathcal{L}_{\text{\tiny{hhB}}}+\mathcal{L}_{\text{\tiny{hBB}}}\ee
with
\bea
&&\mathcal{L}_{\text{\tiny{hh}}}=\frac{1}{4}\left[ h_{\m\n}\Box h^{\m\n}-\frac{1}{n-2} h\Box h\right]\nonumber\\
&&\mathcal{L}_{\text{\tiny{BB}}}=\eta^{\a\b}\left[  \frac{1}{n-1}B^\l_{\l\a}B^\t_{\t\b}-B^\l_{\t\a}B^\t_{\l\b}\right]\nonumber\\
&&\mathcal{L}_{\text{\tiny{hhh}}}=\frac{1}{4}h^{\a\b}\left[\partial_\a h_{\m\n}\partial_\b h^{\m\n}-\frac{1}{n-2}\partial_\a h\partial_\b h+2\partial_\l h_{\a\t}\left(\partial^\t h_\b^\l-\partial^\l h_\b^\t\right)+\frac{2}{n-2}\partial_\l h\partial^\l h_{\a\b}\right]\nonumber\\
&&\mathcal{L}_{\text{\tiny{hhB}}}=B^\l_{\m\n}h^{\m\t}\left[\partial_\l h^\n_\t+\partial_\t h^\n_\l-\partial^\n h_{\l\t}+\frac{1}{n-2}\left(\eta_{\l\t}\partial^\n h-\d_\l^\n\partial_\t h\right)\right]\nonumber\\
&&\mathcal{L}_{\text{\tiny{hBB}}}=h^{\a\b}\left[B^\l_{\t\a}B^\t_{\l\b}-  \frac{1}{n-1}B^\l_{\l\a}B^\t_{\t\b}\right]
\eea
 The mixing between the graviton $h_{\m\n}$ and the connection field $B^\l_{\a\b}$ has been eliminated, and besides  the  graviton propagator
\be G_{\mu_1\nu_1\mu_2\nu_2}=-\cfrac{i}{p^2}\left(\eta_{\mu_1\mu_2}\eta_{\nu_1\nu_2}+\eta_{\mu_1\nu_2}\eta_{\nu_1\mu_2}-\eta_{\mu_1\nu_1}\eta_{\mu_2\nu_2}\right)\ee
is the standard one. It is plain that the propagator of the connection field $B^\l_{\a\b}$  is purely algebraic. This setup is then a convenient starting point in order to perform perturbative calculations.

\section{BRST invariance.}
In order to study the gauge invariance of the cubic action, the best strategy is to explicit the nilpotent BRST transformations. Acting on ghost fields
\bea
&s c^\m=-c^\r \pd_\r c^\m\nonumber\\
&s \overline{c}_\m=b_\m\nonumber\\
&s b_\m=0
\eea
then the   transformation on  $\frak{g}^{\m\n}$
\be\label{g} s\frak{g}^{\m\n}=\frak{g}^{\l\n}\partial_\l c^\m+\frak{g}^{\m\l}\partial_\l c^\n-\partial_\l(c^\l\frak{g}^{\m\n})\ee

On the other hand, the transformation of the connection field yields
\be s\Gamma^\m_{\a\b}=\partial_\l c^\m\Gamma^\l_{\a\b}-\partial_\a c^\l\Gamma^\m_{\l\b}-\partial_\b c^\l\Gamma^\m_{\a\l}-\partial_\a\partial_\b c^\m-c^\l\partial_\l\Gamma^\m_{\a\b}\ee
which means that acting on  $A^\l_{\m\n}$
\be sA^\m_{\a\b}=-c^\l \partial_\l A^\m_{\a\b}+\partial_\l c^\m A^\l_{\a\b}-\partial_\a c^\l A^\m_{\l\b}-\partial_\b c^\l A^\m_{\a\l}-\partial_\a\partial_\b c^\m+\frac{1}{2}\d^\m_\a\partial_\b\partial_\l c^\l+\frac{1}{2}\d^\m_\b\partial_\a\partial_\l c^\l\ee

\par
As we have already mentioned, when studying quantum perturbations around flat space, it is convenient to define
\be\frak{g}^{\a\b}=\eta^{\a\b}-h^{\a\b}\ee
so that
\be\label{h} sh^{\a\b}=-\eta^{\l\b}\partial_\l c^\a-\eta^{\a\l}\partial_\l c^\b+\eta^{\a\b}\partial_\l c^\l+h^{\l\b}\partial_\l c^\a+h^{\a\l}\partial_\l c^\b-\partial_\l(c^\l h^{\a\b})\ee

The gauge fixing Lagrangian yields (cf. Appendix \ref{B})
\be
\mathcal{L}_{\text{\tiny{gf}}}=\bar{c}_\m\left(\frac{1}{2}b^\m-\partial_\n  h^{\m\n}\right)\ee
in such a way that
\be
s\mathcal{L}_{\text{\tiny{gf}}}=b_\m\left(\frac{1}{2}b^\m-\partial_\n h^{\m\n}\right)+\bar{c}_\m\partial_\n(sh^{\m\n})\ee
The ghost Lagrangian reads
\be
\mathcal{L}_{\text{\tiny{gh}}}=\bar{c}_\m\partial_\n sh^{\m\n}=-\bar{c}_\m\Box c^\m+\bar{c}_\m\partial_\n \left[h^{\l\m}\partial_\l c^\n+h^{\l\n}\partial_\l c^\m-\partial_\l(c^\l h^{\m\n})\right]\ee

To quantize the theory, we shall use the field $B^\m_{\a\b}$, defined above, rather than the field $A^\m_{\a\b}$. The BRST transformation  of $B^\m_{\a\b}$
reads
\begin{equation}
s B^\m_{\a\b}=s A^\m_{\a\b}-\frac{1}{2}\left(\partial_\m s h^\l_\n+\partial_\n s h^\l_\m-\partial^\l s h_{\m\n}+\frac{1}{n-2}\eta_{\m\n}\partial^\l s h\right).
\label{Bvariation}
\end{equation}
It can be easily shown that on all fields the transformation is indeed nilpotent $s^2=0$.
\par
 The reader should also bear in mind that $s B^\m_{\a\b}$, unlike $s A^{\lambda}_{\mu\nu}$ and $s h^{\mu\nu}$, has got no contribution which depends
only on the field $c^\mu$. A little computation shows this.

\section{Explicit computation of the propagators.}
To carry out the computations presented in this section, we have used the Feynman rules displayed in Appendix \ref{A}. The following definitions are needed to turn those Feynman rules into mathematical expressions:
\begin{equation}
\begin{array}{l}
{\langle h_{\mu_1\nu_1}(p) h_{\mu_2\nu_2}(-p)\rangle_{0}\!=\!-\cfrac{i}{p^2}\left(\eta_{\mu_1\mu_2}\eta_{\nu_1\nu_2}+\eta_{\mu_1\nu_2}\eta_{\nu_1\mu_2}-\eta_{\mu_1\nu_1}\eta_{\mu_2\nu_2}\right),}\\[8pt]
{\langle B^{\lambda_1}_{\mu_1\nu_1}(p) B^{\lambda_2}_{\mu_2\nu_2}(-p)\rangle_{0}
=-\cfrac{i}{2}\Big[\cfrac{1}{2}\,\d^{\lambda_2}_{(\mu_1}\eta_{\nu_1)(\mu_2}\d^{\lambda_1}_{\nu_2)}+}\\[4pt]
{\phantom{\langle A^{\lambda_1}_{\mu_1\nu_1}(p) A^{\lambda_2}_{\mu_2\nu_2}(-p)\rangle_{0}=-\cfrac{i}{2}\Big[\cfrac{1}{2}\quad\quad}
\eta^{\lambda_1\lambda_2}\,\Big(\cfrac{1}{n-2}\eta_{\mu_1\nu_1}\eta_{\mu_2\nu_2}-\cfrac{1}{2}\eta_{\mu_1(\mu_2}\eta_{\nu_2)\nu_1}\Big)\Big],}\\[8pt]
{\langle c^{\mu_1}(p)\bar{c}_{\mu_2}(p)\rangle_{0}\!=\!-\cfrac{i}{p^2}\,\delta^{\mu_1}_{\mu_2},}\\[8pt]
{V_{(hhh)}^{\mu_1\nu_1\mu_2\nu_2\mu_3\nu_3}(p_1,p_2,p_3)=\cfrac{i}{4}\Big[\cfrac{1}{2}\left(\eta^{\m_1(\m_2}\eta^{\n_2)(\m_3}\eta^{\n_3)\n_1}+\eta^{\n_1(\m_2}\eta^{\n_2)(\m_3}\eta^{\n_3)\m_1}\right)p_2^\l p_{3\l}-}\\[4pt]	{\phantom{V_{(hhh)}^{\mu_1\nu_1\mu_2\nu_2\mu_3\nu_3}(p_1,p_2,p_3)}-\cfrac{1}{(n-2)}\left(\eta^{\m_2\n_2}\eta^{\m_1(\m_3}\eta^{\n_3)\n_2}+\eta^{\m_3\n_3}\eta^{\m_1(\m_2}\eta^{\n_2)\n_1}\right)p_2^\l p_{3\l}+}\\[8pt]
{\phantom{V_{(hhh)}^{\mu_1\nu_1\mu_2\nu_2\mu_3\nu_3}(p_1,p_2,p_3)}+\left(\cfrac{1}{n-2}\eta^{\m_2\n_2}\eta^{\m_3\n_3}-\cfrac{1}{2}\eta^{\m_2(\m_3}\eta^{\n_3)\n_2}\right)p_{2}^{(\m_1}p_{3}^{\n_1)}-}\\[8pt]
	{\phantom{V_{(hhh)}^{\mu_1\nu_1\mu_2\nu_2\mu_3\nu_3}(p_1,p_2,p_3)}-\cfrac{1}{2}p_{2}^{(\m_3}\eta^{\n_3)(\m_1}\eta^{\n_1)(\m_2}p_{3}^{\n_2)}\Big]+}\\[8pt]
	{\phantom{V_{(hhh)}^{\mu_1\nu_1\mu_2\nu_2\mu_3\nu_3}(p_1,p_2,p_3)}+\text{\tiny{$\left\{p_1,\,\m_1\n_1\leftrightarrow p_3,\,\m_3\n_3\right\}$}}+\text{\tiny{$\left\{p_1,\,\m_1\n_1\leftrightarrow p_2,\,\m_2\n_2\right\}$}}+\text{\tiny{$\left\{p_3,\,\m_3\n_3\leftrightarrow p_2,\,\m_2\n_2\right\}$}}                           }\\[8pt]
{V_{(Bhh)\,\lambda}^{\phantom{\lambda}\mu_1\nu_1\mu_2\nu_2\mu_3\nu_3}(p_1,p_2,p_3)=\cfrac{1}{4} \Big[-\cfrac{1}{2}\left(\eta^{\m_1(\m_2}\eta^{\n_2)(\m_3}\eta^{\n_3)\n_1}+\eta^{\n_1(\m_2}\eta^{\n_2)(\m_3}\eta^{\n_3)\m_1}\right)p_{1\l}+}\\[4pt]
	{\phantom{V_{(Bhh)\,\lambda}^{\phantom{\lambda}\mu_1\nu_1\mu_2\nu_2\mu_3\nu_3}(p_1,p_2,p_3)}+\cfrac{1}{2}\d_\l^{(\m_2}\eta^{\n_2)(\n_1}\eta^{\m_1)(\m_3}p_{2}^{\n_3)}-\cfrac{1}{2}\d_\l^{(\m_2}\eta^{\n_2)(\m_3}\eta^{\n_3)(\m_1}p_{2}^{\n_1)}+}\\[8pt]
	{\phantom{V_{(Bhh)\,\lambda}^{\phantom{\lambda}\mu_1\nu_1\mu_2\nu_2\mu_3\nu_3}(p_1,p_2,p_3)}+\cfrac{1}{n-2}\eta^{\m_2\n_2}\d_\l^{(\n_3}\eta^{\m_3)(\m_1}p_{2}^{\n_1)}-\cfrac{1}{n-2}\eta^{\m_2\n_2}\d_\l^{(\m_1}\eta^{\n_1)(\m_3}p_{2}^{\n_3)}\Big]+}\\[8pt]
		{\phantom{V_{(Bhh)\,\lambda}^{\phantom{\lambda}\mu_1\nu_1\mu_2\nu_2\mu_3\nu_3}(p_1,p_2,p_3)}+\text{\tiny{$\left\{p_2,\,\m_2\n_2\leftrightarrow p_3,\,\m_3\n_3\right\}$}}  }\\[4pt]
{V_{(BBh)\,\lambda_1\lambda_2}^{\phantom{\lambda}\mu_1\nu_1\mu_2\nu_2\mu_3\nu_3}(p_1,p_2,p_3)=\cfrac{i}{4}\left[\d_{\l_1}^{(\n_2}\eta^{\m_2)(\m_3}\eta^{\n_3)(\m_1}\d_{\l_2}^{\n_1)}-\cfrac{1}{n-1}\d_{\l_1}^{(\m_1}\eta^{\n_1)(\m_3}\eta^{\n_3)(\n_2}\d_{\l_2}^{\m_2)}\right]   }\\[4pt]
{V_{(\bar{c}ch)\,\mu_2}^{\mu_1\mu_3\nu_3}(p_1,p_2,p_3)=i\cfrac{\kappa}{2}\,\Big(p_{1\mu_2}p_{2}^{(\mu_3}\eta^{\nu_3 )\mu_1}+p_1^{(\mu_3}p_2^{\nu_3 )}\d_{\mu_1}^{\mu_2}
-p_{2\mu_2}p_1^{(\mu_3}\eta^{\nu_3)\mu_1}}\\[8pt]
{\phantom{V_{(\bar{c}c h)\,\mu_2}^{\mu_1\mu_3\nu_3}(p_1,p_2,p_3)=i\cfrac{\kappa}{2}\,\,p_{1\mu_2}
p_{1}^{(\nu_3}\eta^{\mu_3 )\mu_2}+p_2^{(\mu_2}p_1^{\nu_2)}\d_{\mu_1}^{\mu_2}}
\quad\quad\quad\quad -p_{3\mu_2}p_1^{(\mu_3}\eta^{\nu_3 )\mu_1}\Big),}\\[4pt]
{V^{(\text{RHO})\,\mu\nu\mu_1\nu_1}_{\mu_2}(p_1,p_2)=-\cfrac{\kappa}{2}\,\Big[p_2^{\nu_1}\left(\eta^{\mu\mu_1}\d^\nu_{\mu_2}+\eta^{\nu\mu_1}\d^\mu_{\mu_2}\right)+
p_2^{\mu_1}\left(\eta^{\mu\nu_1}\d^{\nu}_{\mu_2}+\eta^{\nu\nu_1}\d^\mu_{\mu_2}\right)}\\[4pt]
{\phantom{V^{(\text{RHO})\,\mu\nu\mu_1\nu_1}_{\mu_2}(p_1,p_2)=-\cfrac{\kappa}{2}\,\Big[p_2^{\nu_1}(\eta^{\mu\mu_1}\d^\nu_{\mu_2}+\eta}
-(p_1+p_2)_{\mu_2} (\eta^{\mu\mu_1}\eta^{\nu\nu_1}+\eta^{\mu\nu_1}\eta^{\nu\mu_1})\Big].}
\end{array}
\label{Feynsymbols}
\end{equation}

 Before we display the results that we have obtained, let us remark that to do the computations that lead to those results we have used the symbolic manipulation softwares FORM \cite{Ruijl:2017dtg} and Mathematica  \cite{mathematica}.

\subsection{The renormalized graviton propagator at one-loop and its Slavnov-Taylor identity}

The one-loop contribution to the graviton propagator is obtained attaching legs to its 1PI part, $\Gamma^{(hh)}_{\mu\nu\rho\sigma}(p)$. This 1PI function is given by
\begin{equation}
\Gamma^{(hh)}_{\mu\nu\rho\sigma}(p)=\Gamma^{(1a)}_{\mu\nu\rho\sigma}(p)+\Gamma^{(1b)}_{\mu\nu\rho\sigma}(p)+\Gamma^{(1c)}_{\mu\nu\rho\sigma}(p)+\Gamma^{(1d)}_{\mu\nu\rho\sigma}(p)
\label{1PIhhsum}
\end{equation}
where $\Gamma^{(1a)}_{\mu\nu\rho\sigma}(p)$, $\Gamma^{(1b)}_{\mu\nu\rho\sigma}(p)$,  $\Gamma^{(1c)}_{\mu\nu\rho\sigma}(p)$ and $\Gamma^{(1d)}_{\mu\nu\rho\sigma}(p)$ are, respectively, the 1PI contributions coming from the Feynman diagrams in Figure 1. To regularize these Feynman diagrams, we shall use dimensional regularization. The regularization parameter $\epsilon$ is defined in terms of the spacetime dimension $n$ in dimensional regularization by the formula: $n=4+2\epsilon$. We have obtained the following results:
\begin{equation*}
\begin{array}{l}
{\Gamma^{(1a)}_{\mu\nu\rho\sigma}(p)=\,\sum_{i=1}^5\,C_i^{(1a)}\;{\cal T}^{(i)}_{\mu\nu\rho\sigma}(p)}\\[8pt]
{\Gamma^{(1b)}_{\mu\nu\rho\sigma}(p)=\,\sum_{i=1}^5\,C_i^{(1b)}\;{\cal T}^{(i)}_{\mu\nu\rho\sigma}(p)}
\end{array}
\end{equation*}
where
\begin{equation*}
\begin{array}{l}
{{\cal T}^{(1)}_{\mu\nu\rho\sigma}(p)=(p^2)^2\,(\eta_{\mu\rho}\eta_{\nu\sigma}+\eta_{\mu\sigma}\eta_{\nu\rho}),\quad
{\cal T}^{(2)}_{\mu\nu\rho\sigma}(p)=(p^2)^2\,\eta_{\mu\nu}\eta_{\rho\sigma},}\\[8pt]
{{\cal T}^{(3)}_{\mu\nu\rho\sigma}(p)=p^2\,(p_\mu p_\nu\, \eta_{\rho\sigma}+p_\rho p_\sigma\,\eta_{\mu\nu}),}\\[8pt]
{{\cal T}^{(4)}_{\mu\nu\rho\sigma}(p)=p^2\,(p_\mu p_\rho\, \eta_{\nu\sigma}+p_\nu p_\sigma\, \eta_{\mu\rho}+p_\mu p_\sigma\, \eta_{\nu\rho}+p_\nu p_\rho\, \eta_{\mu\sigma}),}\\[8pt]
{{\cal T}^{(5)}_{\mu\nu\rho\sigma}(p)=p_\mu p_\nu p_\rho p_\sigma},
\end{array}
\end{equation*}
and
\begin{equation*}
\begin{array}{l}
{C_1^{(1a)}=i\cfrac{\kappa^2}{16\pi^2}\,\Big(-\cfrac{17}{48\epsilon}-\cfrac{17}{48}\ln\Big[-\cfrac{e^\gamma\,p^2}{4\pi\mu^2}\,\Big]+\cfrac{161}{288}\,\Big)+O(\epsilon),}\\[8pt]
{C_2^{(1a)}=i\cfrac{\kappa^2}{16\pi^2}\,\Big(\cfrac{11}{48\epsilon}+\cfrac{11}{48}\ln\Big[ -\cfrac{e^\gamma\,p^2}{4\pi\mu^2}\,\Big]-\cfrac{119}{288}\,\Big)+O(\epsilon),}\\[8pt]
{C_3^{(1a)}=i\cfrac{\kappa^2}{16\pi^2}\,\Big(-\cfrac{13}{24\epsilon}-\cfrac{13}{24}\ln\Big[ -\cfrac{e^\gamma\,p^2}{4\pi\mu^2}\,\Big]+\cfrac{29}{18}\,\Big)+O(\epsilon),}\\[8pt]
{C_4^{(1a)}=i\cfrac{\kappa^2}{16\pi^2}\,\Big(\cfrac{1}{3\epsilon}+\cfrac{1}{3}\ln\Big[ -\cfrac{e^\gamma\,p^2}{4\pi\mu^2}\,\Big]-\cfrac{145}{288}\,\Big)+O(\epsilon),}\\[8pt]
{C_5^{(1a)}=i\cfrac{\kappa^2}{16\pi^2}\,\Big(-\cfrac{11}{6\epsilon}-\cfrac{11}{6}\ln\Big[ -\cfrac{e^\gamma\,p^2}{4\pi\mu^2}\,\Big]+\cfrac{32}{9}\,\Big)+O(\epsilon),}
\end{array}
\end{equation*}
and
\begin{equation*}
\begin{array}{l}
{C_1^{(1b)}=i\cfrac{\kappa^2}{16\pi^2}\,\Big(\cfrac{1}{60\epsilon}+\cfrac{1}{60}\ln\Big[ -\cfrac{e^\gamma\,p^2}{4\pi\mu^2}\,\Big]-\cfrac{77}{1800}\,\Big)+O(\epsilon),}\\[8pt]
{C_2^{(1b)}=C_1^{(1b)},}\\[8pt]
{C_3^{(1b)}=i\cfrac{\kappa^2}{16\pi^2}\,\Big(\cfrac{13}{120\epsilon}+\cfrac{13}{120}\ln\Big[-\cfrac{e^\gamma\,p^2}{4\pi\mu^2}\,\Big]-\cfrac{56}{225}\,\Big)+O(\epsilon),}\\[8pt]
{C_4^{(1b)}=i\cfrac{\kappa^2}{16\pi^2}\,\Big(\cfrac{1}{240\epsilon}+\cfrac{1}{240}\ln\Big[ -\cfrac{e^\gamma\,p^2}{4\pi\mu^2}\,\Big]-\cfrac{23}{1800}\,\Big)+O(\epsilon),}\\[8pt]
{C_5^{(1b)}=i\cfrac{\kappa^2}{16\pi^2}\,\Big(\cfrac{7}{15\epsilon}+\cfrac{7}{15}\ln\Big[ -\cfrac{e^\gamma\,p^2}{4\pi\mu^2}\,\Big]-\cfrac{157}{225}\,\Big)+O(\epsilon),}
\end{array}
\end{equation*}
The fact that the propagator of the $B^{\lambda}_{\mu\nu}$ --see Figure 5 in Appendix \ref{A} and (\ref{Feynsymbols})-- has no momentum dependent denominator immediately leads to the conclusion that
\begin{equation*}
\Gamma^{(1c)}_{\mu\nu\rho\sigma}(p)=0=\Gamma^{(1d)}_{\mu\nu\rho\sigma}(p),
\end{equation*}
for they involve only dimensionally regularized tadpole-like integrals.
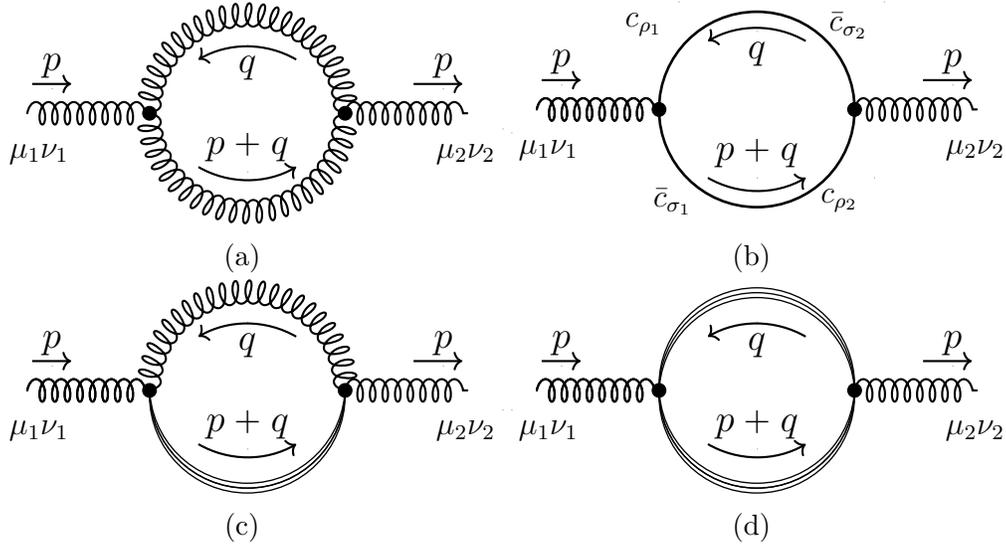
\begin{figure}
	\centering
	\begin{subfigure}[b]{0.4\textwidth}
		\begin{tikzpicture}[scale=1.3,
		decoration={coil,amplitude=4.25,segment length=4.75},anchor=base, baseline]
		\draw[decorate,line width=0.8,black] (0,0) arc (0:180:1);
		\draw[decorate,line width=0.8,black] (-2,0) arc (-180:0:1);
		\draw[spring] (-3.25,0) -- (-2,0);
		\draw[spring] (0,0) -- (1.25,0);
		\filldraw [black] (0,0) circle (2pt) node[anchor=south]{};
		\filldraw [black] (-2,0) circle (2pt) node[anchor=south]{};
		\draw[thick, <-,black](1.2,0.3) -- (0.7,0.3);
		\draw[thick, ->,black](-0.5,0.55) arc (60:120:1);
		\draw[thick, <-,black](-0.5,-0.55) arc (-60:-120:1);
		\filldraw [black] (-1,0.7) circle (0pt) node[anchor=north]{\large$q$};
		\filldraw [black] (-1,-0.6) circle (0pt) node[anchor=south]{\large$p+q$};
		\filldraw [black] (-3.6,-0.2) circle (0pt) node[anchor=north west]{$\mu_1\n_1 $};
		\filldraw [black] (1.7,-0.2) circle (0pt) node[anchor=north east]{$\mu_2\n_2 $};
		\filldraw [black] (-3,0.25) circle (0pt) node[anchor=south]{\large$p$};
		\draw[thick, ->,black](-3.2,0.3) -- (-2.8,0.3);
		\filldraw [black] (1,0.25) circle (0pt) node[anchor=south]{\large$p$};
		\end{tikzpicture}
		\caption{}\label{fig:3DG1}
	\end{subfigure}
 \begin{subfigure}[b]{0.4\textwidth}
	\begin{tikzpicture}[scale=1.3,
	decoration={coil,amplitude=4.25,segment length=4.75},anchor=base, baseline]
	
	\draw[spring] (-3.25,0) -- (-2,0);
	\draw[line width=1] (0,0) arc (0:180:1);
	\draw[line width=1] (-2,0) arc (-180:0:1);
	\draw[spring] (0,0) -- (1.25,0);
	\filldraw [black] (-2,0) circle (2pt) node[anchor=south]{};
	\filldraw [black] (0,0) circle (2pt) node[anchor=south]{};
	
	\draw[thick, <-,black](1.2,0.3) -- (0.7,0.3);
	\draw[thick, ->,black](-0.5,0.7) arc (60:120:1);
	\draw[thick, <-,black](-0.5,-0.7) arc (-60:-120:1);
	\draw[spring] (-3.25,0) -- (-2,0);
	\filldraw [black] (-1,0.8) circle (0pt) node[anchor=north]{\large$q$};
	\filldraw [black] (-1,-0.7) circle (0pt) node[anchor=south]{\large$p+q$};
	\filldraw [black] (-3.6,-0.2) circle (0pt) node[anchor=north west]{$\mu_1\n_1 $};
	\filldraw [black] (1.7,-0.2) circle (0pt) node[anchor=north east]{$\mu_2\n_2 $};
	\filldraw [black] (0.3,1.1) circle (0pt) node[anchor=north east]{$\bar{c}_{\s_2} $};
	\filldraw [black] (-1.8,1.1) circle (0pt) node[anchor=north east]{$c_{\r_1}$};
	\filldraw [black] (0.2,-0.7) circle (0pt) node[anchor=north east]{$c_{\r_2} $};
	\filldraw [black] (-1.5,-0.7) circle (0pt) node[anchor=north east]{$\bar{c}_{\s_1} $};
	\filldraw [black] (-3,0.25) circle (0pt) node[anchor=south]{\large$p$};
	\draw[thick, ->,black](-3.2,0.3) -- (-2.8,0.3);
	\filldraw [black] (1,0.25) circle (0pt) node[anchor=south]{\large$p$};

	\end{tikzpicture}
	\caption{}\label{fig:3DG2}
\end{subfigure}
\begin{subfigure}[b]{0.4\textwidth}
	\begin{tikzpicture}[scale=1.3,
	decoration={coil,amplitude=4.25,segment length=4.75},anchor=base, baseline]
	
	\draw[spring] (-3.25,0) -- (-2,0);
		\draw[decorate,line width=0.8,black] (0,0) arc (0:180:1);
	\draw[line width=.5] (-2,-0.05) arc (-180:0:1);
	\draw[line width=.5] (-2,0) arc (-180:0:1);
	\draw[line width=.5] (-2,0.05) arc (-180:0:1);
	\draw[line width=0.5] (-2,0) arc (-180:0:1);
	\draw[spring] (0,0) -- (1.25,0);
	\filldraw [black] (-2,0) circle (2pt) node[anchor=south]{};
	\filldraw [black] (0,0) circle (2pt) node[anchor=south]{};
	
	\draw[thick, <-,black](1.2,0.3) -- (0.7,0.3);
	\draw[thick, ->,black](-0.5,0.55) arc (60:120:1);
	\draw[thick, <-,black](-0.5,-0.55) arc (-60:-120:1);
	\draw[spring] (-3.25,0) -- (-2,0);
	\filldraw [black] (-1,0.7) circle (0pt) node[anchor=north]{\large$q$};
	\filldraw [black] (-1,-0.6) circle (0pt) node[anchor=south]{\large$p+q$};
	\filldraw [black] (-3.6,-0.2) circle (0pt) node[anchor=north west]{$\mu_1\n_1 $};
	\filldraw [black] (1.7,-0.2) circle (0pt) node[anchor=north east]{$\mu_2\n_2 $};
	\filldraw [black] (-3,0.25) circle (0pt) node[anchor=south]{\large$p$};
	\draw[thick, ->,black](-3.2,0.3) -- (-2.8,0.3);
	\filldraw [black] (1,0.25) circle (0pt) node[anchor=south]{\large$p$};

	\end{tikzpicture}
	\caption{}\label{fig:3DG3}
\end{subfigure}
\begin{subfigure}[b]{0.4\textwidth}
	\begin{tikzpicture}[scale=1.3,
	decoration={coil,amplitude=4.25,segment length=4.75},anchor=base, baseline]
	
	\draw[spring] (-3.25,0) -- (-2,0);
	\draw[line width=0.5] (0,-0.05) arc (0:180:1);
	\draw[line width=0.5] (0,0) arc (0:180:1);
	\draw[line width=0.5] (0,0.05) arc (0:180:1);
	\draw[line width=.5] (-2,-0.05) arc (-180:0:1);
	\draw[line width=.5] (-2,0) arc (-180:0:1);
	\draw[line width=.5] (-2,0.05) arc (-180:0:1);
	\draw[line width=0.5] (-2,0) arc (-180:0:1);
	\draw[spring] (0,0) -- (1.25,0);
	\filldraw [black] (-2,0) circle (2pt) node[anchor=south]{};
	\filldraw [black] (0,0) circle (2pt) node[anchor=south]{};
	
	\draw[thick, <-,black](1.2,0.3) -- (0.7,0.3);
	\draw[thick, ->,black](-0.5,0.55) arc (60:120:1);
	\draw[thick, <-,black](-0.5,-0.55) arc (-60:-120:1);
	\draw[spring] (-3.25,0) -- (-2,0);
	\filldraw [black] (-1,0.7) circle (0pt) node[anchor=north]{\large$q$};
	\filldraw [black] (-1,-0.6) circle (0pt) node[anchor=south]{\large$p+q$};
	\filldraw [black] (-3.6,-0.2) circle (0pt) node[anchor=north west]{$\mu_1\n_1 $};
	\filldraw [black] (1.7,-0.2) circle (0pt) node[anchor=north east]{$\mu_2\n_2 $};
	\filldraw [black] (-3,0.25) circle (0pt) node[anchor=south]{\large$p$};
	\draw[thick, ->,black](-3.2,0.3) -- (-2.8,0.3);
	\filldraw [black] (1,0.25) circle (0pt) node[anchor=south]{\large$p$};

	\end{tikzpicture}
	\caption{}\label{fig:3DG4}
\end{subfigure}
	\caption{Graviton propagator 1-loop diagrams}\label{fig:3}
\end{figure}

Taking into account equation (\ref{1PIhhsum}) and the results displayed below this equation, one concludes that
\begin{equation}
\Gamma^{(hh)}_{\mu\nu\rho\sigma}(p)=\,\sum_{i=1}^5\,C_i^{(hh)}\;{\cal T}^{(i)}_{\mu\nu\rho\sigma}(p),
\label{1PIhh}
\end{equation}
with
\begin{equation}
\begin{array}{l}
{C_1^{(hh)}=i\cfrac{\kappa^2}{16\pi^2}\,\Big(-\cfrac{27}{80\epsilon}-\cfrac{27}{80}\ln\Big[-\cfrac{e^\gamma\,p^2}{4\pi\mu^2}\,\Big]+\cfrac{413}{800}\,\Big)+O(\epsilon),}\\[8pt]
{C_2^{(hh)}=i\cfrac{\kappa^2}{16\pi^2}\,\Big(\cfrac{59}{240\epsilon}+\cfrac{59}{240}\ln\Big[-\cfrac{e^\gamma\,p^2}{4\pi\mu^2}\,\Big]-\cfrac{3283}{7200}\,\Big)+O(\epsilon),}\\[8pt]
{C_3^{(hh)}=i\cfrac{\kappa^2}{16\pi^2}\,\Big(-\cfrac{13}{30\epsilon}-\cfrac{13}{30}\ln\Big[-\cfrac{e^\gamma\,p^2}{4\pi\mu^2}\,\Big]+\cfrac{613}{450}\,\Big)+O(\epsilon),}\\[8pt]
{C_4^{(hh)}=i\cfrac{\kappa^2}{16\pi^2}\,\Big(\cfrac{27}{80\epsilon}+\cfrac{27}{80}\ln\Big[ -\cfrac{e^\gamma\,p^2}{4\pi\mu^2}\,\Big]-\cfrac{413}{800}\,\Big)+O(\epsilon),}\\[8pt]
{C_5^{(hh)}=i\cfrac{\kappa^2}{16\pi^2}\,\Big(-\cfrac{41}{30\epsilon}-\cfrac{41}{30}\ln\Big[ -\cfrac{e^\gamma\,p^2}{4\pi\mu^2}\,\Big]+\cfrac{643}{225}\,\Big)+O(\epsilon).}
\end{array}
\label{Chh}
\end{equation}

Let us now check that $\Gamma^{(hh)}_{\mu\nu\rho\sigma}(p)$ in (\ref{1PIhh}) and (\ref{Chh}) satisfies the corresponding dimensionally regularized Slavnov-Taylor identitity. As shown in Appendix \ref{B}, the Slavnov-Taylor identity in question reads
\begin{equation}
\begin{array}{l}
{i\Big[p^\mu\eta^{\lambda\nu}+p^\nu\eta^{\lambda\mu}-p^\lambda\eta^{\mu\nu}\Big] \Gamma^{(hh)}_{\mu\nu\rho\sigma}(p)=
\cfrac{1}{4}\Big[p^2(\eta_{\mu\rho} \eta_{\nu\sigma}+\eta_{\mu\sigma}\eta_{\nu\rho}}\\[8pt]
{\quad\quad\quad\quad-\cfrac{2}{n-2}\eta_{\rho\sigma}\eta_{\mu\nu})-
(p_\rho p_\mu \eta_{\sigma\nu}+p_\rho p_\nu \eta_{\sigma\mu}+p_\sigma p_\mu \eta_{\rho\nu}+p_\sigma p_\nu \eta_{\rho\mu})\Big] \Gamma^{(\text{RHO})\mu\nu}_{\lambda}(p).}
\end{array}
\label{SThh}
\end{equation}
It is crucial to bear in mind that the $\eta_{\mu\nu}$ and momenta in the previous equation live in the $n$-dimensional spacetime of dimensional regularization as defined in \cite{Breitenlohner:1977hr}. Indeed,  (\ref{SThh}) is a dimensionally regularized identity. Notice  that when computing  either side of (\ref{SThh}) contractions of the type
$\eta^{\mu\nu}\eta_{\mu\nu}=4+2\epsilon$ will occur. These contractions will yield  UV finite contributions when multiplied by $1/\epsilon$; contributions that would be absent
if the $\eta_{\mu\nu}$'s lived in 4 dimensions, this absence leading to a false breaking of the regularized Slavnov-Taylor identity.

Now, $\Gamma^{(\text{RHO})\mu\nu}_{\lambda}(p)$ in  equation (\ref{SThh}) is equal to the value of the Feynman diagram in Figure 2. Thus, we have
\begin{equation}
\begin{array}{l}
{\Gamma^{(\text{RHO})\mu\nu}_{\lambda}(p)=C_1^{(\text{RHO})}\,p^2 p_\lambda\eta^{\mu\nu}+C_2^{(\text{RHO})}\,p^2(p^\mu\delta_\lambda^\nu+p^\nu\delta_\lambda^\mu)+
C_3^{(\text{RHO})}\,p_\lambda p^\mu p^\nu,}\\[8pt]
{C_1^{(\text{RHO})}=\cfrac{\kappa^2}{16\pi^2}\,\Big(-\cfrac{5}{4\epsilon}-\cfrac{5}{4}\ln\Big[-\cfrac{e^\gamma\,p^2}{4\pi\mu^2}\,\Big]+\cfrac{5}{2}\,\Big)+O(\epsilon),}\\[8pt]
{C_2^{(\text{RHO})}=\cfrac{\kappa^2}{16\pi^2}\,\Big(\cfrac{1}{12\epsilon}+\cfrac{1}{12}\ln\Big[-\cfrac{e^\gamma\,p^2}{4\pi\mu^2}\,\Big]-\cfrac{2}{9}\,\Big)+O(\epsilon),}\\[8pt]
{C_3^{(\text{RHO})}=\cfrac{\kappa^2}{16\pi^2}\,\Big(\cfrac{4}{3\epsilon}+\cfrac{4}{3}\ln\Big[-\cfrac{e^\gamma\,p^2}{4\pi\mu^2}\,\Big]-\cfrac{23}{9}\,\Big)+O(\epsilon),}\\[8pt]
\end{array}
\label{1PIRHO}
\end{equation}
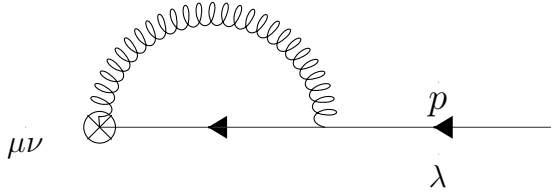
\begin{figure}
	\centering
\begin{subfigure}[b]{0.4\textwidth}
	\begin{tikzpicture}[scale=3,
	decoration={coil,amplitude=4.25,segment length=4.75},anchor=base, baseline]
	\draw[line width=0.8,black,  postaction={decorate, decoration={markings, mark=at position 1 with {\arrow[fill=black, scale=1.5]{Triangle}} } }](0.5,0)-- (0.48,0);
	\draw[line width=0.4,blue!7!black!80] (0.5,0) -- (2,0);
	\draw[line width=0.4,blue!7!black!80](0,0) -- (0.48,0);
	\draw[line width=0.8,black,  postaction={decorate, decoration={markings, mark=at position 1 with {\arrow[fill=black, scale=1.5]{Triangle}} } }](1.5,0)-- (1.48,0);
	\draw[decorate,line width=0.4,black] (1,0) arc (0:180:0.5);
	\filldraw [black] (1.5,-.12) circle (0pt) node[anchor=north ]{$\l$};
	\filldraw [black] (1.5,.2) circle (0pt) node[anchor=north ]{\large$p$};
	\filldraw [black] (-.33,0) circle (0pt) node[anchor=north ]{$\mu\nu$};
	\draw (0,0) circle (2pt);
	\draw (0,0) node[cross=5pt,rotate=90]{};
	\end{tikzpicture}
\end{subfigure}
\caption{Diagram with one  $\r_{\m\n}$ insertion}\label{fig:4}
\end{figure}

Substituting (\ref{1PIhh}) in (\ref{SThh}), one gets the following value for the left hand side of (\ref{SThh}):
\begin{equation*}
\Big(\cfrac{\kappa^2}{16\pi^2}\Big)p^2\Big[p^2 p^\lambda\eta_{\rho\sigma}\,\Big(\cfrac{1}{4\epsilon}+\cfrac{1}{4}\ln\Big[-\cfrac{e^\gamma\,p^2}{4\pi\mu^2}\,\Big]-\cfrac{3}{4}\,\Big)+
p^\lambda p_\rho p_\sigma\,\Big(\cfrac{1}{2\epsilon}+\cfrac{1}{2}\ln\Big[-\cfrac{e^\gamma\,p^2}{4\pi\mu^2}\,\Big]-1\,\Big)\Big]+O(\epsilon).
\end{equation*}
This is the very same result that it is obtained by substituting (\ref{1PIRHO}) in the right hand side of (\ref{SThh}). We thus conclude that the regularized expressions in
(\ref{1PIhh}) and (\ref{1PIRHO}) are consistent with BRST invariance.

We have seen that there is not a clash between the dimensionally regularized theory and  BRST symmetry. But what about the renormalized theory?
It has long been known that when the Ward identities of the dimensionally regularized theory have coefficients which explicitly depends on $n$ --see \cite{Breitenlohner:1977hr}-- and/or involve the trace of $\eta_{\mu\nu}$ --see \cite{Duff:1993wm}-- there is no guarantee that the MS renormalization scheme will yield renormalized Green functions which do  satisfy the correct Ward identities. Indeed, anomalies arise when the  Ward identities do not hold for the MS renormalized theory and  no UV finite counterterms can be added to restore the symmetry broken by the MS renormalization scheme. It is plain that the previous comments are relevant to the theory analyzed in this paper.

 Let us now show that, although the MS renormalization scheme applied to $\Gamma^{(hh)}_{\mu\nu\rho\sigma}(p)$ in (\ref{1PIhh}) and $\Gamma^{(\text{RHO})\lambda}_{\mu\nu}(p)$
 in (\ref{1PIRHO}) yields renormalized functions which do not satisfy --for n=4-- the Slavnov-Taylor identity in (\ref{SThh}), UV finite counterterms can be introduced so that
 the Slavnov-Taylor identity in question is restored. By using  $\Gamma^{(hh)}_{\mu\nu\rho\sigma}(p)$ in (\ref{1PIhh}) and $\Gamma^{(\text{RHO})\lambda}_{\mu\nu}(p)$
 in (\ref{1PIRHO}), we define the following renormalized objects:
 \begin{equation}
\Gamma^{(hh,\,\text{ren})}_{\mu\nu\rho\sigma}(p)=\,\sum_{i=1}^5\,C_i^{(hh,\,\text{ren})}\;{\cal T}^{(i)}_{\mu\nu\rho\sigma}(p),
\label{1PIhhren}
\end{equation}
where
\begin{equation*}
\begin{array}{l}
{C_1^{(hh,\,\text{ren})}=i\cfrac{\kappa^2}{16\pi^2}\,\Big(-\cfrac{27}{80}\ln\Big[-\cfrac{e^\gamma\,p^2}{4\pi\mu^2}\,\Big]+\cfrac{413}{800}+ F_1\,\Big),}\\[8pt]
{C_2^{(hh,\,\text{ren})}=i\cfrac{\kappa^2}{16\pi^2}\,\Big(+\cfrac{59}{240}\ln\Big[-\cfrac{e^\gamma\,p^2}{4\pi\mu^2}\,\Big]-\cfrac{3283}{7200}+F_2\,\Big),}\\[8pt]
{C_3^{(hh,\,\text{ren})}=i\cfrac{\kappa^2}{16\pi^2}\,\Big(-\cfrac{13}{30}\ln\Big[-\cfrac{e^\gamma\,p^2}{4\pi\mu^2}\,\Big]+\cfrac{613}{450}+ F_3\,\Big),}\\[8pt]
{C_4^{(hh,\,\text{ren})}=i\cfrac{\kappa^2}{16\pi^2}\,\Big(+\cfrac{27}{80}\ln\Big[ -\cfrac{e^\gamma\,p^2}{4\pi\mu^2}\,\Big]-\cfrac{413}{800}+F_4\,\Big),}\\[8pt]
{C_5^{(hh,\,\text{ren})}=i\cfrac{\kappa^2}{16\pi^2}\,\Big(-\cfrac{41}{30}\ln\Big[ -\cfrac{e^\gamma\,p^2}{4\pi\mu^2}\,\Big]+\cfrac{643}{225}+F_5\,\Big),}
\end{array}
\end{equation*}
and
\begin{equation}
\Gamma^{(\text{RHO},\text{ren})\mu\nu}_{\lambda}(p)=C_1^{(\text{RHO},\text{ren})}p^2 p_\lambda\eta^{\mu\nu}+C_2^{(\text{RHO},\text{ren})}p^2(p^\mu\delta_\lambda^\nu+p^\nu\delta_\lambda^\mu)+
C_3^{(\text{RHO},\text{ren})}p_\lambda p^\mu p^\nu,
\label{1PIRHOren}
\end{equation}
with
\begin{equation*}
\begin{array}{l}
{C_1^{(\text{RHO},\text{ren})}\!\!=\!\!\cfrac{\kappa^2}{16\pi^2}\,\Big(-\cfrac{5}{4}\ln\Big[-\cfrac{e^\gamma\,p^2}{4\pi\mu^2}\,\Big]+\cfrac{5}{2}+G_1\,\Big),}\\[8pt]
{C_2^{(\text{RHO},\text{ren})}\!\!=\!\!\cfrac{\kappa^2}{16\pi^2}\,\Big(+\cfrac{1}{12}\ln\Big[-\cfrac{e^\gamma\,p^2}{4\pi\mu^2}\,\Big]-\cfrac{2}{9}+G_2\,\Big),}\\[8pt]
{C_3^{(\text{RHO},\text{ren})}\!\!=\!\!\cfrac{\kappa^2}{16\pi^2}\,\Big(+\cfrac{4}{3}\ln\Big[-\cfrac{e^\gamma\,p^2}{4\pi\mu^2}\,\Big]-\cfrac{23}{9}+G_3\,\Big).}
\end{array}
\end{equation*}
The $F_i$, $i=1...5$, and $G_i$, $i=1,2,3$, are constants which must bet chosen so that, for $n=4$, the Slavnov-Taylor identity in (\ref{SThh}) holds for $\Gamma^{(hh,\,\text{ren})}_{\mu\nu\rho\sigma}(p)$ and $\Gamma^{(\text{RHO},\,\text{ren})\mu\nu}_{\lambda}(p)$ above.

It can be shown that for $\Gamma^{(hh,\,\text{ren})}_{\mu\nu\rho\sigma}(p)$ in (\ref{1PIhhren}) and $n=4$ the left hand side of (\ref{SThh}) is equal to
\begin{equation}
\begin{array}{l}
{\Big(\cfrac{\kappa^2}{16\pi^2}\Big)\,\Big[p^2 p^\lambda p_\rho p_\sigma\Big(-\cfrac{2}{15}-F_5+2 F_3+\cfrac{1}{2}\,\ln\Big[-\cfrac{e^\gamma\,p^2}{4\pi\mu^2}\Big]\Big)+}\\[8pt]
{(p^2)^2 (p_\rho\,\delta^\lambda_\sigma+p_\sigma\,\delta^\lambda_\rho)\Big(-2 F_4- 2 F_1\Big)+}\\[8pt]
{(p^2)^2 p^\lambda\eta_{\rho\sigma}\Big(-\cfrac{149}{120}-F_3+2 F_2+ 2 F_1+\cfrac{1}{4} \ln\Big[-\cfrac{e^\gamma\,p^2}{4\pi\mu^2}\Big]\Big)\Big].}
\end{array}
\label{LHS}
\end{equation}

On the other hand, the substitution of $\Gamma^{(\text{RHO},\,\text{ren})\,\mu\nu}_{\lambda}(p)$ in the right hand side of (\ref{SThh}) yields
\begin{equation}
\begin{array}{l}
{\Big(\cfrac{\kappa^2}{16\pi^2}\Big)\,\Big[p^2 p^\lambda p_\rho p_\sigma\Big(-1+\cfrac{1}{2}\,G_3+G_2+G_1+\cfrac{1}{2}\,\ln\Big[-\cfrac{e^\gamma\,p^2}{4\pi\mu^2}\Big]\Big)+}\\[8pt]
{(p^2)^2 p^\lambda \eta_{\rho\sigma}\Big(-\cfrac{1}{2}+\cfrac{1}{4}\,G_ 3+\cfrac{1}{2}\,G_2+\cfrac{1}{2}\,G_1+\cfrac{1}{4} \ln\Big[-\cfrac{e^\gamma\,p^2}{4\pi\mu^2}\Big]\Big)\Big].}
\end{array}
\label{RHS}
\end{equation}
Since (\ref{LHS}) must be equal to (\ref{RHS}) for the Slavnov-Taylor identity to hold, one concludes that $F_i$, $i=1...5$ and $G_i$, $i=1,2,3$ must satisfy the following
equations
\begin{equation}
\begin{array}{l}
{F_4=-F_1,}\\[8pt]
{-\cfrac{2}{15}-F_5+2 F_3=-1+\cfrac{1}{2}\,G_3+G_2+G_1,}\\[8pt]
{-\cfrac{149}{120}-F_3+2 F_2+ 2 F_1=-\cfrac{1}{2}+\cfrac{1}{4}\,G_ 3+\cfrac{1}{2}\,G_2+\cfrac{1}{2}\,G_1}.
\end{array}
\end{equation}
Notice that setting all the $F_i$'s and $G_i$'s to zero --i.e., choosing the MS renormalization scheme-- would not do, but there are plenty of other choices that restore
Slavnov-Taylor identity broken by the MS renormalization scheme. Hence, no anomaly arises. Also notice that setting all the $F_i$'s to zero (regardless of the values of the $G_i$) is not an admissible choice either.
Let us finally make the comment that the clash between the MS renormalization scheme and the BRST symmetry we have just analyzed arises because the pole parts of
$\Gamma^{(hh)}_{\mu\nu\rho\sigma}(p)$ and $\Gamma^{(\text{RHO})\mu\nu}_{\lambda}(p)$ do not satisfy (\ref{SThh}) in $n$ dimensions. Indeed, using only those pole parts, one obtains
\begin{equation*}
\begin{array}{l}
{\Big(\cfrac{\kappa^2}{16\pi^2}\Big)p^2\Big[p^2 p^\lambda\eta_{\rho\sigma}\,\Big(\cfrac{1}{4\epsilon}+\cfrac{59}{120}\,\Big)+
p^\lambda p_\rho p_\sigma\,\Big(\cfrac{1}{2\epsilon}-\cfrac{13}{15}\,\Big)\Big]+O(\epsilon),}\\[8pt]
{\Big(\cfrac{\kappa^2}{16\pi^2}\Big)p^2\Big[p^2 p^\lambda\eta_{\rho\sigma}\,\Big(\cfrac{1}{4\epsilon}-\cfrac{1}{4}\,\Big)+
p^\lambda p_\rho p_\sigma\,\Big(\cfrac{1}{2\epsilon}\Big)\Big]+O(\epsilon),}
\end{array}
\end{equation*}
for the L.H.S and R.H.S of (\ref{SThh}), respectively.
\par
It is interesting  to compare our results with those obtained from the second order Hilbert Lagrangian \cite{Capper}, where also the corresponding  Slavnov-Taylor's identities have been shown to hold.

\subsection{The one-loop  $h_{\rho\sigma}-B^{\lambda}_{\mu\nu}$ propagator}
We have seen that  no contribution to the propagator
\begin{equation}
\left\langle 0\left|T B^\l_{\r\s}(\xp) h_{\m\n}(x)\right| 0\right\rangle
\label{Bhprop}
\end{equation}
occurs at tree level. Here we shall show that such contribution is generated at one loop.

Let us introduce the following set of tensors
\begin{equation}
\begin{array}{l}
{\mathfrak{t}^{\l_2}_{1\,\m_1\n_1\m_2\n_2}(p)= p^2p^{\lambda_2}\eta_{\mu_1\nu_1}\eta_{\mu_2\nu_2},}\\[8pt]
{\mathfrak{t}^{\l_2}_{2\,\m_1\n_1\m_2\n_2}(p)= p^2p^{\lambda_2}\left(\eta_{\mu_1\mu_2}\eta_{\nu_1\nu_2}+\eta_{\mu_1\nu_2}\eta_{\nu_1\mu_2}\right),}\\[8pt]
{\mathfrak{t}^{\l_2}_{3\,\m_1\n_1\m_2\n_2}(p)=p^2\left(\d^{\lambda_2}_{\mu_2}\eta_{\nu_2\mu_1}p_{\nu_1}+\d^{\lambda_2}_{\nu_2}\eta_{\mu_2\mu_1}p_{\nu_1}
+\d^{\lambda_2}_{\mu_2}\eta_{\nu_2\nu_1}p_{\mu_1}+\d^{\lambda_2}_{\nu_2}\eta_{\mu_2\nu_1}p_{\mu_1}\right),}\\[8pt]
{\mathfrak{t}^{\l_2}_{4\,\m_1\n_1\m_2\n_2}(p)=p^2\left(\d^{\lambda_2}_{\mu_1}\eta_{\nu_1\nu_2}p_{\mu_2}+\d^{\lambda_2}_{\nu_1}\eta_{\mu_1\nu_2}p_{\mu_2}
+\d^{\lambda_2}_{\mu_1}\eta_{\nu_1\mu_2}p_{\nu_2}+\d^{\lambda_2}_{\nu_1}\eta_{\mu_1\mu_2}p_{\nu_2}\right),}\\[8pt]
{\mathfrak{t}^{\l_2}_{5\,\m_1\n_1\m_2\n_2}(p)= p^2\eta_{\mu_1\nu_1}\left(\d^{\lambda_2}_{\mu_2} p_{\nu_2}+\d^{\lambda_2}_{\nu_2} p_{\mu_2}\right),}\\[8pt]
{\mathfrak{t}^{\l_2}_{6\,\m_1\n_1\m_2\n_2}(p)= p^{\lambda_2} p_{\mu_1} p_{\nu_1}\eta_{\mu_2\nu_2},}\\[8pt]
{\mathfrak{t}^{\l_2}_{7\,\m_1\n_1\m_2\n_2}(p)= p^{\lambda_2}\left(p_{\mu_1} p_{\mu_2}\eta_{\nu_1\nu_2}+p_{\nu_1} p_{\mu_2}\eta_{\mu_1\nu_2}+p_{\mu_1} p_{nu_2}\eta_{\nu_1\mu_2}+p_{\nu_1} p_{\nu_2}\eta_{\mu_1\mu_2}\right),}\\[8pt]
{\mathfrak{t}^{\l_2}_{8\,\m_1\n_1\m_2\n_2}(p)= p^{\lambda_2} p_{\mu_2} p_{\nu_2}\eta_{\mu_1\nu_1},}\\[8pt]
{\mathfrak{t}^{\l_2}_{9\,\m_1\n_1\m_2\n_2}(p)= p_{\mu_1} p_{\nu_1}\left(\d^{\lambda_2}_{\mu_2} p_{\nu_2}+\d^{\lambda_2}_{\nu_2} p_{\mu_2}\right),}\\[8pt]
{\mathfrak{t}^{\l_2}_{10\,\m_1\n_1\m_2\n_2}(p)= p_{\mu_2} p_{\nu_2}\left(\d^{\lambda_2}_{\mu_1} p_{\nu_1}+\d^{\lambda_2}_{\n_1}p_{\mu_1}\right).}\\[8pt]
\end{array}
\end{equation}
Then, the one-loop contribution to the propagator in (\ref{Bhprop}) is obtained from the one-loop 1PI function, say $\Gamma^{\l_2}_{\m_1\n_1\m_2\n_2}(p)$,
\begin{equation*}
	\Gamma^{\lambda_2}_{\m_1\n_1\m_2\n_2}(p)=	\Gamma^{(3a)\lambda_2}_{\m_1\n_1\m_2\n_2}(p)+	\Gamma^{(3b)\lambda_2}_{\m_1\n_1\m_2\n_2}(p)
\end{equation*}
where $	\Gamma^{(3a)\lambda_2}_{\m_1\n_1\m_2\n_2}(p)$ and  $\Gamma^{(3b)\lambda_2}_{\m_1\n_1\m_2\n_2}(p)$are, respectively, the 1PI contributions coming from the Feynman diagrams in Figure 3.

By working out the Feynman integrals we have obtained the following  results
\begin{equation*}
	\begin{array}{l}
{\Gamma^{(3a)\lambda_2}_{\m_1\n_1\m_2\n_2}(p)=\sum_{i=1}^{10}\,E_{i}\,\mathfrak{t}^{\lambda_2}_{i\,\m_1\n_1\m_2\n_2}(p)}\\[8pt]
{\Gamma^{(3b)\lambda_2}_{\m_1\n_1\m_2\n_2}(p)=0,}\\[8pt]
\end{array}
\end{equation*}
with
\begin{equation*}
\begin{array}{l}
{E_1=-\cfrac{\kappa^2}{16\pi^2}\,\Big(\cfrac{13}{24\epsilon}+\cfrac{13}{24}\ln\Big[-\cfrac{e^\gamma\,p^2}{4\pi\mu^2}\,\Big]-\cfrac{113}{72}\,\Big)+O(\epsilon),}\\[8pt]
{E_2=-\cfrac{\kappa^2}{16\pi^2}\,\Big(-\cfrac{1}{2\epsilon}-\cfrac{1}{2}\ln\Big[ -\cfrac{e^\gamma\,p^2}{4\pi\mu^2}\,\Big]+\cfrac{2}{3}\,\Big)+O(\epsilon),}\\[8pt]
{E_3=-\cfrac{\kappa^2}{16\pi^2}\,\Big(-\cfrac{1}{48\epsilon}-\cfrac{1}{48}\ln\Big[ -\cfrac{e^\gamma\,p^2}{4\pi\mu^2}\,\Big]+\cfrac{1}{18}\,\Big)+O(\epsilon),}\\[8pt]
{E_4=-\cfrac{\kappa^2}{16\pi^2}\,\Big(\cfrac{19}{48\epsilon}+\cfrac{19}{48}\ln\Big[ -\cfrac{e^\gamma\,p^2}{4\pi\mu^2}\,\Big]-\cfrac{49}{72}\,\Big)+O(\epsilon),}\\[8pt]
{E_5=-\cfrac{\kappa^2}{16\pi^2}\,\Big(-\cfrac{3}{8\epsilon}-\cfrac{3}{8}\ln\Big[ -\cfrac{e^\gamma\,p^2}{4\pi\mu^2}\,\Big]+\cfrac{13}{12}\,\Big)+O(\epsilon),}\\[8pt]
{E_6=-\cfrac{\kappa^2}{16\pi^2}\,\Big(\cfrac{5}{6\epsilon}+\cfrac{5}{6}\ln\Big[-\cfrac{e^\gamma\,p^2}{4\pi\mu^2}\,\Big]-\cfrac{37}{18}\,\Big)+O(\epsilon),}\\[8pt]
{E_7=-\cfrac{\kappa^2}{16\pi^2}\,\Big(\cfrac{1}{4\epsilon}+\cfrac{1}{4}\ln\Big[ -\cfrac{e^\gamma\,p^2}{4\pi\mu^2}\,\Big]-\cfrac{1}{3}\,\Big)+O(\epsilon),}\\[8pt]
{E_8=-\cfrac{\kappa^2}{16\pi^2}\,\Big(\cfrac{1}{3\epsilon}+\cfrac{1}{3}\ln\Big[ -\cfrac{e^\gamma\,p^2}{4\pi\mu^2}\,\Big]-\cfrac{19}{18}\,\Big)+O(\epsilon),}\\[8pt]
{E_9=-\cfrac{\kappa^2}{16\pi^2}\,\Big(\cfrac{1}{12\epsilon}+\cfrac{1}{12}\ln\Big[ -\cfrac{e^\gamma\,p^2}{4\pi\mu^2}\,\Big]-\cfrac{1}{18}\,\Big)+O(\epsilon),}\\[8pt]
{E_{10}=-\cfrac{\kappa^2}{16\pi^2}\,\Big(-\cfrac{3}{4\epsilon}-\cfrac{3}{4}\ln\Big[ -\cfrac{e^\gamma\,p^2}{4\pi\mu^2}\,\Big]+\cfrac{5}{4}\,\Big)+O(\epsilon).}\\[8pt]
\end{array}
\end{equation*}
It is plain that the one-loop contribution we have obtained cannot be removed by a local redefinition of the fields.

\begin{figure}
	\centering
\begin{subfigure}[b]{0.4\textwidth}
	\begin{tikzpicture}[scale=1.3,
	decoration={coil,amplitude=4.25,segment length=4.75},anchor=base, baseline]
	\draw[decorate,line width=0.8,black] (0,0) arc (0:180:1);
	\draw[decorate,line width=0.8,black] (-2,0) arc (-180:0:1);
	\draw[spring] (-3.25,0) -- (-2,0);
	\draw[thick, ->,black](-0.5,0.55) arc (60:120:1);
	\draw[thick, <-,black](-0.5,-0.55) arc (-60:-120:1);
\filldraw [black] (0,0) circle (2pt) node[anchor=south]{};
\filldraw [black] (-2,0) circle (2pt) node[anchor=south]{};
	\filldraw [black] (-1,0.7) circle (0pt) node[anchor=north]{\large$q$};
	\filldraw [black] (-1,-0.6) circle (0pt) node[anchor=south]{\large$p+q$};
	\filldraw [black] (-3.2,-0.2) circle (0pt) node[anchor=north west]{$\mu_1\n_1 $};
	\filldraw [black] (-2.5,0.25) circle (0pt) node[anchor=south]{\large$p$};
	\draw[thick, ->,black](-2.7,0.3) -- (-2.3,0.3);
	\draw[line width=0.5] (0,-0.05) -- (1.25,-0.05);
	\draw[line width=0.5] (0,0) -- (1.25,0);
	\draw[line width=0.5] (0,0.05) -- (1.25,0.05);
	
	\filldraw [black] (1.7,-0.2) circle (0pt) node[anchor=north east]{$\mu_2\n_2 $};
	\filldraw [black] (1.6,0.6) circle (0pt) node[anchor=north east]{$\l_2 $};
	\filldraw [black] (0.6,.25) circle (0pt) node[anchor=south]{\large$p$};
	\draw[thick, <-,black](0.8,0.3) -- (0.4,0.3);
	\end{tikzpicture}
	\caption{}\label{fig:4DG1}
\end{subfigure}
\begin{subfigure}[b]{0.4\textwidth}
	\begin{tikzpicture}[scale=1.3,
	decoration={coil,amplitude=4.25,segment length=4.75},anchor=base, baseline]
	\draw[decorate,line width=0.8,black] (0,0) arc (0:180:1);
	\draw[line width=.5] (-2,-0.05) arc (-180:0:1);
	\draw[line width=.5] (-2,0) arc (-180:0:1);
	\draw[line width=.5] (-2,0.05) arc (-180:0:1);
	\draw[spring] (-3.25,0) -- (-2,0);
	\draw[thick, ->,black](-0.5,0.55) arc (60:120:1);
	\draw[thick, <-,black](-0.5,-0.55) arc (-60:-120:1);
	\filldraw [black] (0,0) circle (2pt) node[anchor=south]{};
	\filldraw [black] (-2,0) circle (2pt) node[anchor=south]{};
	\filldraw [black] (-1,0.7) circle (0pt) node[anchor=north]{\large$q$};
	\filldraw [black] (-1,-0.6) circle (0pt) node[anchor=south]{\large$p+q$};
	\filldraw [black] (-3.2,-0.2) circle (0pt) node[anchor=north west]{$\mu_1\n_1 $};
	\filldraw [black] (-2.5,0.25) circle (0pt) node[anchor=south]{\large$p$};
	\draw[thick, ->,black](-2.7,0.3) -- (-2.3,0.3);
	\draw[line width=0.5] (0,-0.05) -- (1.25,-0.05);
	\draw[line width=0.5] (0,0) -- (1.25,0);
	\draw[line width=0.5] (0,0.05) -- (1.25,0.05);
	
	\filldraw [black] (1.7,-0.2) circle (0pt) node[anchor=north east]{$\mu_2\n_2 $};
	\filldraw [black] (1.6,0.6) circle (0pt) node[anchor=north east]{$\l_2 $};
	\filldraw [black] (0.6,.25) circle (0pt) node[anchor=south]{\large$p$};
	\draw[thick, <-,black](0.8,0.3) -- (0.4,0.3);
	\end{tikzpicture}
	\caption{}\label{fig:4DG2}
\end{subfigure}
	\caption{Graviton-B propagator 1-loop diagram}\label{fig:4}
\end{figure}

\subsection{The one-loop  $B^{\lambda}_{\mu\nu}-B^{\lambda}_{\mu\nu}$ propagator}

The one-loop contribution to the propagator of the $B^{\lambda}_{\mu\nu}$  can be retrieved by using its 1PI momentum contribution $\Gamma^{\lambda_1\lambda_2}_{\mu_1\nu_1\mu_2\nu_2}(p)$. In the case at hand, this 1PI function is given by the sum
\begin{equation}
 \Gamma^{\lambda_1\lambda_2}_{\mu_1\nu_1\mu_2\nu_2}(p)=\Gamma^{(4a)\,\lambda_1\lambda_2}_{\mu_1\nu_1\mu_2\nu_2}(p)+\Gamma^{(4b)\,\lambda_1\lambda_2}_{\mu_1\nu_1\mu_2\nu_2}(p),
\label{1PIBB}
\end{equation}
where $\Gamma^{(4a)\,\lambda_1\lambda_2}_{\mu_1\nu_1\mu_2\nu_2}(p)$ and $\Gamma^{(4b)\,\lambda_1\lambda_2}_{\mu_1\nu_1\mu_2\nu_2}(p)$ are given by the 1PI Feynman diagrams in Figure 4, respectively

Since the Feynman diagram b) in Figure 4 only involves tadpole-like integrals, one concludes that
\begin{equation}
\Gamma^{(4b)\,\lambda_1\lambda_2}_{\mu_1\nu_1\mu_2\nu_2}(p)=0,
\label{Gamma4b}
\end{equation}
in dimensional regularization.

Let us define the following set of tensors:
\begin{equation}
\begin{array}{l}
{{\cal T}^{\lambda_1\lambda_2}_{1\,\mu_1\nu_1\mu_2\nu_2}(p)=p^2\eta^{\lambda_1\lambda_{2}}\eta_{\mu_{1}\nu_{1}}\eta_{\mu_{2}\nu_{2}},}\\[8pt]
{{\cal T}^{\lambda_1\lambda_2}_{2\,\mu_1\nu_1\mu_2\nu_2}(p)=p^2\eta^{\lambda_1\lambda_{2}}\left(\eta_{\mu_{1}\nu_{2}}\eta_{\nu_{1}\mu_{2}}+
\eta_{\mu_{1}\mu_{2}}\eta_{\nu_{1}\nu_{2}}\right),}\\[8pt]
{{\cal T}^{\lambda_1\lambda_2}_{3\,\mu_1\nu_1\mu_2\nu_2}(p)=p^2\left[\eta_{\mu_{1}\nu_{1}}\left(\delta^{\lambda_1}_{\mu_2}\delta^{\lambda_2}_{\nu_2}
+\delta^{\lambda_1}_{\nu_2}\delta^{\lambda_2}_{\mu_2}\right)+
\eta_{\mu_{2}\nu_{2}}\left(\delta^{\lambda_1}_{\mu_1}\delta^{\lambda_2}_{\nu_1}+\delta^{\lambda_1}_{\nu_1}\delta^{\lambda_2}_{\mu_1}\right)\right],}\\[8pt]
{{\cal T}^{\lambda_1\lambda_2}_{4\,\mu_1\nu_1\mu_2\nu_2}(p)=p^2\left(\delta^{\lambda_1}_{\mu_1}\delta^{\lambda_2}_{\mu_2}\eta_{\nu_{1}\nu_{2}}
+\delta^{\lambda_1}_{\nu_1}\delta^{\lambda_2}_{\mu_2}\eta_{\mu_{1}\nu_{2}}
+\delta^{\lambda_1}_{\mu_1}\delta^{\lambda_2}_{\nu_2}\eta_{\nu_{1}\mu_{2}}+\delta^{\lambda_1}_{\nu_1}\delta^{\lambda_2}_{\nu_2}\eta_{\mu_{1}\mu_{2}}\right),}\\[8pt]
{{\cal T}^{\lambda_1\lambda_2}_{5\,\mu_1\nu_1\mu_2\nu_2}(p) =p^2\left(\delta^{\lambda_1}_{\mu_2}\delta^{\lambda_2}_{\mu_1}\eta_{\nu_{1}\nu_{2}}+\delta^{\lambda_1}_{\mu_2}\delta^{\lambda_2}_{\nu_1}\eta_{\mu_{1}\nu_{2}}
+\delta^{\lambda_1}_{\nu_2}\delta^{\lambda_2}_{\mu_1}\eta_{\nu_{1}\mu_{2}}+\delta^{\lambda_1}_{\nu_2}\delta^{\lambda_2}_{\nu_1}\eta_{\mu_{1}\mu_{2}}\right),}\\[8pt]
{{\cal T}^{\lambda_1\lambda_2}_{6\,\mu_1\nu_1\mu_2\nu_2}(p)=p^{\lambda_1} p^{\lambda_2}\eta_{\mu_{1}\nu_{1}}\eta_{\mu_{2}\nu_{2}},}\\[8pt]
{{\cal T}^{\lambda_1\lambda_2}_{7\,\mu_1\nu_1\mu_2\nu_2}(p)=p^{\lambda_1} p^{\lambda_2}\left(\eta_{\mu_{1}\nu_{2}}\eta_{\nu_{1}\mu_{2}}+\eta_{\mu_{1}\mu_{2}}\eta_{\nu_{1}\nu_{2}}\right),}\\[8pt]
{{\cal T}^{\lambda_1\lambda_2}_{8\,\mu_1\nu_1\mu_2\nu_2}(p)=\eta_{\mu_{1}\nu_{1}}\left(\d^{\lambda_1}_{\mu_2} p^{\lambda_2} p_{\nu_2}+\d^{\lambda_1}_{\nu_2} p^{\lambda_2} p_{\mu_2}\right)+\eta_{\mu_{2}\nu_{2}}\left(\d^{\lambda_2}_{\mu_1} p^{\lambda_1} p_{\nu_1}+\d^{\lambda_2}_{\nu_1} p^{\lambda_1} p_{\mu_1}\right)),}\\[8pt]
{{\cal T}^{\lambda_1\lambda_2}_{9\,\mu_1\nu_1\mu_2\nu_2}(p)=\eta_{\mu_{1}\mu_{2}}\left(\d^{\lambda_1}_{\nu_1} p^{\lambda_2} p_{\nu_2}+\d^{\lambda_2}_{\nu_2} p^{\lambda_1} p_{\nu_1}\right)+\eta_{\nu_{1}\mu_{2}}\left(\d^{\lambda_1}_{\mu_1} p^{\lambda_2} p_{\nu_2}+\d^{\lambda_2}_{\nu_2} p^{\lambda_1} p_{\mu_1}\right)}\\[8pt]
{\quad\quad\quad+\eta_{\mu_{1}\nu_{2}}\left(\d^{\lambda_1}_{\nu_1} p^{\lambda_2} p_{\mu_2}+\d^{\lambda_2}_{\mu_2} p^{\lambda_1} p_{\nu_1}\right)+\eta_{\nu_{1}\nu_{2}}\left(\d^{\lambda_1}_{\mu_1} p^{\lambda_2} p_{\mu_2}+\d^{\lambda_2}_{\mu_2} p^{\lambda_1} p_{\mu_1}\right),}\\[8pt]
{{\cal T}^{\lambda_1\lambda_2}_{10\,\mu_1\nu_1\mu_2\nu_2}(p)=\eta_{\mu_{1}\mu_{2}}\left(\d^{\lambda_1}_{\nu_2} p^{\lambda_2} p_{\nu_1}+\d^{\lambda_2}_{\nu_1} p^{\lambda_1} p_{\nu_2}\right)+\eta_{\nu_{1}\mu_{2}}\left(\d^{\lambda_1}_{\nu_2} p^{\lambda_2} p_{\mu_1}+\d^{\lambda_2}_{\mu_1} p^{\lambda_1} p_{\nu_2}\right)+}\\[8pt]
{\quad\quad\quad+\eta_{\mu_{1}\nu_{2}}\left(\d^{\lambda_1}_{\mu_2} p^{\lambda_2} p_{\nu_1}+\d^{\lambda_2}_{\nu_1} p^{\lambda_1} p_{\mu_2}\right)+\eta_{\nu_{1}\nu_{2}}\left(\d^{\lambda_1}_{\mu_2} p^{\lambda_2} p_{\mu_1}+\d^{\lambda_2}_{\mu_1} p^{\lambda_1} p_{\mu_2}\right),}\\[8pt]
{{\cal T}^{\lambda_1\lambda_2}_{11\,\mu_1\nu_1\mu_2\nu_2}(p)=\eta_{\mu_{1}\nu_{1}}\left(\d^{\lambda_2}_{\nu_2} p^{\lambda_1} p_{\mu_2}+\d^{\lambda_2}_{\mu_2} p^{\lambda_1} p_{\nu_2}\right)+\eta_{\mu_{2}\nu_{2}}\left(\d^{\lambda_1}_{\mu_1} p^{\lambda_2} p_{\nu_1}+\d^{\lambda_1}_{\nu_1} p^{\lambda_2} p_{\mu_1}\right),}\\[8pt]
{{\cal T}^{\lambda_1\lambda_2}_{12\,\mu_1\nu_1\mu_2\nu_2}(p)=\eta^{\lambda_{1}\lambda_{2}}\left(\eta_{\mu_{1}\nu_{1}}p_{\mu_2} p_{\nu_2}+\eta_{\mu_{2}\nu_{2}} p_{\mu_1} p_{\nu_1}\right),}\\[8pt]
{{\cal T}^{\lambda_1\lambda_2}_{13\,\mu_1\nu_1\mu_2\nu_2}(p)=\eta^{\lambda_{1}\lambda_{2}}\left(\eta_{\mu_{1}\mu_{2}}p_{\nu_1} p_{\nu_2}+\eta_{\nu_{1}\mu_{2}}p_{\mu_1} p_{\nu_2}+\eta_{\mu_{1}\nu_{2}}p_{\nu_1} p_{\mu_2}+\eta_{\nu_{1}\nu_{2}}p_{\mu_1} p_{\mu_2}\right),}\\[8pt]
{{\cal T}^{\lambda_1\lambda_2}_{14\,\mu_1\nu_1\mu_2\nu_2}(p)=p_{\mu_1} p_{\nu_1}\left(\d^{\lambda_1}_{\mu_2}\d^{\lambda_2}_{\nu_2}+\d^{\lambda_1}_{\nu_2}\d^{\lambda_2}_{\mu_2}\right)+p_{\mu_2} p_{\nu_2}\left(\d^{\lambda_1}_{\mu_1}\d^{\lambda_2}_{\nu_1}+\d^{\lambda_1}_{\nu_1}\d^{\lambda_2}_{\mu_1}\right),}\\[8pt]
{{\cal T}^{\lambda_1\lambda_2}_{15\,\mu_1\nu_1\mu_2\nu_2}(p)=p_{\mu_1} p_{\mu_2} \d^{\lambda_1}_{\nu_1}\d^{\lambda_2}_{\nu_2}+p_{\mu_1} p_{\nu_2} \d^{\lambda_1}_{\nu_1}\d^{\lambda_2}_{\mu_2}+p_{\nu_1} p_{\mu_2} \d^{\lambda_1}_{\mu_1}
\d^{\lambda_2}_{\nu_2}+p_{\nu_1} p_{\nu_2} \d^{\lambda_1}_{\mu_1}\d^{\lambda_2}_{\mu_2},}\\[8pt]
{{\cal T}^{\lambda_1\lambda_2}_{16\,\mu_1\nu_1\mu_2\nu_2}(p)=p_{\mu_1} p_{\mu_2} \d^{\lambda_1}_{\nu_2}\d^{\lambda_2}_{\nu_1}+p_{\mu_1} p_{\nu_2} \d^{\lambda_1}_{\mu_2}\d^{\lambda_2}_{\nu_1}+p_{\nu_1} p_{\mu_2} \d^{\lambda_1}_{\nu_2}\d^{\lambda_2}_{\mu_1}+
p_{\nu_1} p_{\nu_2} \d^{\lambda_1}_{\mu_2}\d^{\lambda_2}_{\mu_1}).}
\end{array}
\end{equation}

Then, some computations lead to the following result:
\begin{equation}
\Gamma^{(4a)\,\lambda_1\lambda_2}_{\mu_1\nu_1\mu_2\nu_2}(p)=\sum_{i=1}^{16}\,C_i^{(4a)}\,{\cal T}^{\lambda_1\lambda_2}_{i\,\mu_1\nu_1\mu_2\nu_2}(p),
\label{Gamma4a}
\end{equation}
where
\begin{equation*}
\begin{array}{l}
{C_1^{(4a)}=i\cfrac{\kappa^2}{16\pi^2}\,\Big(-\cfrac{1}{6\epsilon}-\cfrac{1}{6}\ln\Big[-\cfrac{e^\gamma\,p^2}{4\pi\mu^2}\,\Big]+\cfrac{4}{9}\,\Big)+O(\epsilon),}\\[8pt]
{C_2^{(4a)}=i\cfrac{\kappa^2}{16\pi^2}\,\Big(\cfrac{1}{8\epsilon}+\cfrac{1}{8}\ln\Big[ -\cfrac{e^\gamma\,p^2}{4\pi\mu^2}\,\Big]-\cfrac{5}{24}\,\Big)+O(\epsilon),}\\[8pt]
{C_3^{(4a)}=i\cfrac{\kappa^2}{16\pi^2}\,\Big(\cfrac{1}{24\epsilon}+\cfrac{1}{24}\ln\Big[ -\cfrac{e^\gamma\,p^2}{4\pi\mu^2}\,\Big]-\cfrac{11}{72}\,\Big)+O(\epsilon),}\\[8pt]
{C_4^{(4a)}=i\cfrac{\kappa^2}{16\pi^2}\,\Big(\cfrac{1}{16\epsilon}+\cfrac{1}{16}\ln\Big[ -\cfrac{e^\gamma\,p^2}{4\pi\mu^2}\,\Big]-\cfrac{3}{16}\,\Big)+O(\epsilon),}\\[8pt]
{C_5^{(4a)}=i\cfrac{\kappa^2}{16\pi^2}\,\Big(-\cfrac{7}{48\epsilon}-\cfrac{7}{48}\ln\Big[ -\cfrac{e^\gamma\,p^2}{4\pi\mu^2}\,\Big]+\cfrac{53}{144}\,\Big)+O(\epsilon),}\\[8pt]
{C_6^{(4a)}=i\cfrac{\kappa^2}{16\pi^2}\,\Big(-\cfrac{1}{\epsilon}-\ln\Big[-\cfrac{e^\gamma\,p^2}{4\pi\mu^2}\,\Big]+2\,\Big)+O(\epsilon),}\\[8pt]
{C_7^{(4a)}=i\cfrac{\kappa^2}{16\pi^2}\,\Big(-\cfrac{13}{12\epsilon}-\cfrac{13}{12}\ln\Big[ -\cfrac{e^\gamma\,p^2}{4\pi\mu^2}\,\Big]+\cfrac{59}{36}\,\Big)+O(\epsilon),}\\[8pt]
{C_8^{(4a)}=i\cfrac{\kappa^2}{16\pi^2}\,\Big(-\cfrac{1}{6\epsilon}-\cfrac{1}{6}\ln\Big[ -\cfrac{e^\gamma\,p^2}{4\pi\mu^2}\,\Big]+\cfrac{13}{36}\,\Big)+O(\epsilon),}\\[8pt]
{C_9^{(4a)}=i\cfrac{\kappa^2}{16\pi^2}\,\Big(-\cfrac{1}{6\epsilon}-\cfrac{1}{6}\ln\Big[ -\cfrac{e^\gamma\,p^2}{4\pi\mu^2}\,\Big]+\cfrac{4}{9}\,\Big)+O(\epsilon),}\\[8pt]
{C_{10}^{(4a)}=i\cfrac{\kappa^2}{16\pi^2}\,\Big(\cfrac{17}{24\epsilon}+\cfrac{17}{24}\ln\Big[ -\cfrac{e^\gamma\,p^2}{4\pi\mu^2}\,\Big]-\cfrac{85}{72}\,\Big)+O(\epsilon),}\\[8pt]
{C_{11}^{(4a)}=i\cfrac{\kappa^2}{16\pi^2}\,\Big(\cfrac{1}{4\epsilon}+\cfrac{1}{4}\ln\Big[-\cfrac{e^\gamma\,p^2}{4\pi\mu^2}\,\Big]-\cfrac{1}{2}\,\Big)+O(\epsilon),}\\[8pt]
{C_{12}^{(4a)}=i\cfrac{\kappa^2}{16\pi^2}\,\Big(\cfrac{1}{3\epsilon}+\cfrac{1}{3}\ln\Big[ -\cfrac{e^\gamma\,p^2}{4\pi\mu^2}\,\Big]-\cfrac{13}{18}\,\Big)+O(\epsilon),}\\[8pt]
{C_{13}^{(4a)}=i\cfrac{\kappa^2}{16\pi^2}\,\Big(-\cfrac{1}{3\epsilon}-\cfrac{1}{3}\ln\Big[ -\cfrac{e^\gamma\,p^2}{4\pi\mu^2}\,\Big]+\cfrac{17}{36}\,\Big)+O(\epsilon),}\\[8pt]
{C_{14}^{(4a)}=i\cfrac{\kappa^2}{16\pi^2}\,\Big(\cfrac{1}{6}\Big)+O(\epsilon),}\\[8pt]
{C_{15}^{(4a)}=i\cfrac{\kappa^2}{16\pi^2}\,\Big(-\cfrac{1}{24\epsilon}-\cfrac{1}{24}\ln\Big[ -\cfrac{e^\gamma\,p^2}{4\pi\mu^2}\,\Big]-\cfrac{1}{72}\,\Big)+O(\epsilon),}\\[8pt]
{C_{16}^{(4a)}=i\cfrac{\kappa^2}{16\pi^2}\,\Big(-\cfrac{5}{24\epsilon}-\cfrac{5}{24}\ln\Big[-\cfrac{e^\gamma\,p^2}{4\pi\mu^2}\,\Big]+\cfrac{13}{72}\,\Big)+O(\epsilon).}
\end{array}
\end{equation*}

In view of (\ref{1PIBB}), (\ref{Gamma4a}) and (\ref{Gamma4b}), we conclude that
\begin{equation}
\Gamma^{\lambda_1\lambda_2}_{\mu_1\nu_1\mu_2\nu_2}(p)=\Gamma^{(4a)\,\lambda_1\lambda_2}_{\mu_1\nu_1\mu_2\nu_2}(p).
\label{1PIBBres}
\end{equation}

\begin{figure}[htbp]
	\centering
		\begin{subfigure}[b]{0.4\textwidth}
		\begin{tikzpicture}[scale=1.3,
		decoration={coil,amplitude=4.25,segment length=4.75},anchor=base, baseline]
		\draw[line width=0.5] (-3.25,-0.05) -- (-2,-0.05);
		\draw[line width=0.5] (-3.25,0) -- (-2,0);
		\draw[line width=0.5] (-3.25,0.05) -- (-2,0.05);
		\draw[decorate,line width=0.8,black] (0,0) arc (0:180:1);
		\draw[decorate,line width=0.8,black] (-2,0) arc (-180:0:1);
		\draw[line width=0.5] (0,-0.05) -- (1.25,-0.05);
		\draw[line width=0.5] (0,0) -- (1.25,0);
		\draw[line width=0.5] (0,0.05) -- (1.25,0.05);
		\filldraw [black] (-2,0) circle (2pt) node[anchor=south]{};
		\filldraw [black] (0,0) circle (2pt) node[anchor=south]{};
		
		\draw[thick, ->,black](-0.5,0.55) arc (60:120:1);
		\draw[thick, <-,black](-0.5,-0.55) arc (-60:-120:1);
		\filldraw [black] (-1,0.7) circle (0pt) node[anchor=north]{\large$q$};
		\filldraw [black] (-1,-0.6) circle (0pt) node[anchor=south]{\large$p+q$};
		\filldraw [black] (-3.6,-0.2) circle (0pt) node[anchor=north west]{$\mu_1\n_1 $};
		\filldraw [black] (-3.5,0.6) circle (0pt) node[anchor=north west]{$\l_1$};
		\filldraw [black] (1.7,-0.2) circle (0pt) node[anchor=north east]{$\mu_2\n_2 $};
		\filldraw [black] (1.6,0.6) circle (0pt) node[anchor=north east]{$\l_2 $};
		\filldraw [black] (-2.5,0.25) circle (0pt) node[anchor=south]{\large$p$};
		\draw[thick, ->,black](-2.7,0.3) -- (-2.3,0.3);
		\filldraw [black] (0.6,.25) circle (0pt) node[anchor=south]{\large$p$};
		\draw[thick, <-,black](0.8,0.3) -- (0.4,0.3);

		\end{tikzpicture}
		\caption{}\label{fig:6DG1}
	\end{subfigure}
	\begin{subfigure}[b]{0.4\textwidth}
		\begin{tikzpicture}[scale=1.3,
		decoration={coil,amplitude=4.25,segment length=4.75},anchor=base, baseline]
		\draw[line width=0.5] (-3.25,-0.05) -- (-2,-0.05);
		\draw[line width=0.5] (-3.25,0) -- (-2,0);
		\draw[line width=0.5] (-3.25,0.05) -- (-2,0.05);
		\draw[decorate,line width=0.8,black] (0,0) arc (0:180:1);
		\draw[line width=.5] (-2,-0.05) arc (-180:0:1);
		\draw[line width=.5] (-2,0) arc (-180:0:1);
		\draw[line width=.5] (-2,0.05) arc (-180:0:1);
		\draw[line width=0.5] (0,-0.05) -- (1.25,-0.05);
		\draw[line width=0.5] (0,0) -- (1.25,0);
		\draw[line width=0.5] (0,0.05) -- (1.25,0.05);
		\filldraw [black] (-2,0) circle (2pt) node[anchor=south]{};
		\filldraw [black] (0,0) circle (2pt) node[anchor=south]{};
		
		\draw[thick, ->,black](-0.5,0.55) arc (60:120:1);
		\draw[thick, <-,black](-0.5,-0.55) arc (-60:-120:1);
			\filldraw [black] (-1,0.7) circle (0pt) node[anchor=north]{\large$q$};
		\filldraw [black] (-1,-0.6) circle (0pt) node[anchor=south]{\large$p+q$};
		\filldraw [black] (-3.6,-0.2) circle (0pt) node[anchor=north west]{$\mu_1\n_1 $};
		\filldraw [black] (-3.5,0.6) circle (0pt) node[anchor=north west]{$\l_1$};
		\filldraw [black] (1.7,-0.2) circle (0pt) node[anchor=north east]{$\mu_2\n_2 $};
		\filldraw [black] (1.6,0.6) circle (0pt) node[anchor=north east]{$\l_2 $};
		\filldraw [black] (-2.5,0.25) circle (0pt) node[anchor=south]{\large$p$};
		\draw[thick, ->,black](-2.7,0.3) -- (-2.3,0.3);
		\filldraw [black] (0.6,.25) circle (0pt) node[anchor=south]{\large$p$};
		\draw[thick, <-,black](0.8,0.3) -- (0.4,0.3);

		\end{tikzpicture}
		\caption{}\label{fig:6DG2}
\end{subfigure}
\caption{$B^\l_{\m\n}$ propagator 1-loop diagrams}\label{fig:6}
\end{figure}
A final comment. Notice that the counterterms needed to renormalize $\Gamma^{\lambda_1\lambda_2}_{\mu_1\nu_1\mu_2\nu_2}(p)$ in (\ref{1PIBBres}) are not part of the tree level action. Indeed, the theory is not renormalizable by power-counting.
\par
Again, this is similar to what happens when studying quantum fluctuations around flat space in Hilbert's action.

\section{First order versus second order: renormalization of composite operators.}
On the face of it, it seems somewhat disturbing that our results for the different Green functions and in particular for the graviton propagator are quite different from the classical results obtained with a second order formalism, given the fact that we have claimed not only that first order Lagrangians and second order ones are classically equivalent, but also that the background field gauge effective Lagrangians are equivalent at least to one loop order.
\par
We shall  argue in this paragraph  that the propagator for the connection field
\be
\left.\left\langle 0\left|T B^\a_{\b\g}(x) B^\m_{\n\r}(\xp)\right|0 \right\rangle\right|_{FO}
\ee
we have computed in the present paper, corresponds in a second order approach to the Green function of the normal composite operator $B^\a_{\b\g}(x)$ defined by using the equation of motion and Zimmerman's \cite{Zimmermann} approach, $N\left[B^\a_{\b\g}(x)\right]$,  based in Bogoliubov's renormalization techniques.
\be
\left\langle 0\left|T B^\a_{\b\g}(x) B^\m_{\n\r}(\xp)\right|0 \right\rangle\left.\right|_{FO}=\left\langle 0\left|T N[B^\a_{\b\g}(x)]N[ B^\m_{\n\r}(\xp)]\right|0 \right\rangle\left.\right|_{SO}
\ee
Let us focus into  the first order Cheung-Remmen Lagrangian, although the argument holds for every Palatini-like first order Lagrangian
\be \mathcal{L}_{\text{\tiny{CR}}}=\frak{g}^{\a\b}\left(  \frac{1}{n-1}A^\l_{\l\a}A^\t_{\t\b}-A^\l_{\t\a}A^\t_{\l\b}\right)-A^\l_{\a\b}\partial_\l\frak{g}^{\a\b}\ee

The vacuum persistence amplitude (generating functional of Green functions) reads
\be
Z\left[J,j\right]=\int \mathcal{D}\frak{g}\mathcal{D}A~e^{-\text{\tiny{$\frac{i}{2\kappa^2}$}}\int d^nx\left(A\Delta A-A\partial \frak{g}+AJ+\frak{g}j\right)}\ee
where we have defined the kernel
\bea
&&\Delta_{\t~~\l}^{\r\s\m\n}=\frac{1}{n-1}\d^{(\m}_\l\frak{g}^{\n)(\s}\d^{\r)}_\t-\d^{(\m}_\t\frak{g}^{\n)(\s}\d^{\r)}_\l+\text{\tiny{$\left\{\m\n\l\leftrightarrow\r\s\t\right\}$}}\nonumber\\
\eea

Here  $J^\l_{\a\b}(x)$ and $j_{\m\n}(x)$ are the sources of the fields $A^\l_{\a\b}(x)$ and $\frak{g}_{\m\n}(x)$ respectively.
We have suppressed the index in the integral in order not to clutter the formulas too much.
As a matter of fact
\be
\left\langle 0\left|T A(x)A(\xp)\right|0\right\rangle= \left.\frac{\d^2 Z\left[J,j\right]}{\d J(x)\d J(x')}\right|_{J=j=0}
\ee

Completing squares in the   integrand over ${\cal D}A$,
\be A\Delta A+2CA=\left[C+A\Delta\right]G\left[\Delta A+C\right]-CGC\ee
where
\be 2C_\l^{\a\b}=J_\l^{\a\b}-\partial_\l\frak{g}^{\a\b}\ee
and
\be \Delta_{\t~~\g}^{\r\s\m\n}G^{\g~~\l}_{\m\n\a\b}=\d^\l_\t\d^{\r\s}_{\a\b}\ee
 The inverse kernel is  algebraic
\bea
G^{\t~~\l}_{\r\s\m\n}&&=\frac{1}{2}\frak{g}^{\l\t}\left(\frak{g}_{\m\r}\frak{g}_{\n\s}+\frak{g}_{\m\s}\frak{g}_{\n\r}-\frac{2}{n-2}\frak{g}_{\m\n}\frak{g}_{\r\s}\right)-\nonumber\\
&&-\frac{1}{2}\Big[\d^\l_\r\left(\d^\t_\m\frak{g}_{\s\n}+\d^\t_\n\frak{g}_{\s\m}\right)+\d^\l_\s\left(\d^\t_\m\frak{g}_{\r\n}+\d^\t_\n\frak{g}_{\r\m}\right)\Big]\eea
in conclusion
\bea
Z\left[J,j\right]&&=\text{det}G^{-\frac{1}{2}}\int \mathcal{D}\frak{g}~e^{-\text{\tiny{$\frac{i}{2\kappa^2}$}}\int d^nx\left(\frak{g}j-CGC\right)}
=\nonumber\\
&&=\text{det}G^{-\frac{1}{2}}\int \mathcal{D}\frak{g}~e^{-\text{\tiny{$\frac{i}{2\kappa^2}$}}\int d^nx\left(\frak{g}j+\mathcal{L}_{\text{\tiny{H}}}+\frac{1}{2}JG\partial\frak{g}-\frac{1}{4}JGJ\right)}
\eea
because $-\frac{1}{4}\partial\frak{g}G\partial\frak{g}=\mathcal{L}_{\text{\tiny{H}}}$.
 \par
 The classical equation of motion $ \frac{1}{2}G^{\t~~\l}_{\r\s\m\n}\partial_\l\frak{g}^{\m\n}=A^\t_{\r\s}$
conveys the fact that this action can now be interpreted as a { \em second order} Hilbert type action with sources for the composite operator $A^\m_{\r\s}(x)$. Then
\be \left.\frac{\d^2 Z\left[J,j\right]}{\d J(x)\d J(x')}\right|_{J=j=0}=-\frac{1}{2}\langle G\rangle+\frac{1}{4}\langle G\partial\frak{g}G\partial\frak{g}\rangle=\left\langle 0\left|T A(x) A(\xp) \right|0\right\rangle\ee
where the ultralocal term $G$ only contributes when $x=\xp$, id est, when the arguments  of the corresponding composite operators  coincide.

\section{Conclusions}
In this paper some aspects of the cubic action for gravity recently proposed in \cite{Cheung} have been discussed. We have explicitly related it with the general  first order  Palatini approach, and examined in detail the corresponding BRST invariance.
\par
In this theory the two-point function of the field $B^\m_{\n\r}$ is divergent; and besides, the counterterms needed to remove the divergence in question  are not included in the tree level action, as shown by explicit computation. We have worked out the renormalized  graviton propagator $\left\langle 0\left|T h_{\m\n}(x) h_{\r\s}(\xp)\right| 0\right\rangle$ in such a way that the corresponding Slavnov-Taylor identity is fulfilled. Also by explicit computation, we have  shown that the  propagator $\left\langle 0\left|T B^\l_{\r\s}(\xp)h_{\m\n}(x)\right| 0\right\rangle$, which vanishes at tree-level, gets a one-loop radiative correction. This  correction cannot be removed by making a local  field redefinition.
\par
All these seems a bit paradoxical, given the fact that we have claimed that first order and second order actions are equivalent not only classically but also to one loop order in a general background field gauge expansion. In fact
we have argued  that Green's functions of the connection $A^\m_{\n\r}(x)$ field computed with the cubic action
\be
\left.\left\langle 0\left|T A_{\a\b}^\m(x) A^\n_{\r\s}(\xp) \right|0\right\rangle\right|_{FO}
\ee
are  directly related to the normal products of the composite operator $A_c[\fg]$
\be
\left.\left\langle 0\left|T N\left[(A_c)_{\a\b}^\m[\fg]\right ]N\left[ (A_c)^\n_{\r\s}[\fg]\right] \right|0\right\rangle\right|_{SO}
\ee
computed with  the second order Hilbert action and where the composite field $(A_c)^\m_{\n\r}(x)$ is formally defined by plugging the equation of motion for the $A^\m_{\n \r}(x)$ field given in terms of $\fg_{\a\b}(x)$.
 \par

\section{Acknowledgements.}
We should like to thank  E. Velasco-Aja for help with the figures. EA and JA acknowledge partial financial support by the Spanish MINECO through the Centro de excelencia Severo Ochoa Program  under Grant CEX2020-001007-S  funded by MCIN/AEI/10.13039/501100011033. EA and JA also acknowledge partial financial support by the Spanish Research Agency (Agencia Estatal de Investigaci\'on) through the grant PID2022-137127NB-I00 funded by MCIN/AEI/10.13039/501100011033/ FEDER, UE.
EA and JA acknowledge the European Union's Horizon 2020 research and innovation programme under the Marie Sklodowska-Curie grant agreement No 860881-HIDDeN and also byGrant PID2019-108892RB-I00 funded by MCIN/AEI/ 10.13039/501100011033 and by ``ERDF A way of making Europe''.
The work of CPM has been financially  supported by the Spanish Ministry of Science, Innovation and Universities through grant PID2023-149834NB-I00.

\newpage
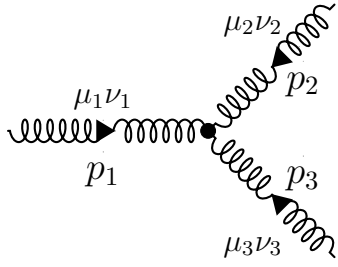
\appendix
\section{Feynman Rules}\label{A}
\begin{figure}[htbp]
	\begin{subfigure}[b]{0.4\textwidth}
		\begin{tikzpicture}[scale=2.8,
		decoration={coil,amplitude=4.25,segment length=4.75},anchor=base, baseline]
		\draw[line width=0.8,black,  postaction={decorate, decoration={markings, mark=at position 1 with {\arrow[fill=black, scale=1.5]{Triangle}} } }] (-0.54,0)--(-0.541,0);
		\draw[spring] (-.532,0) -- (-1.3,0);
		\draw[spring] (-.47,0) -- (.3,0);
		\filldraw [black] (-1,-0.25) circle (0pt) node[anchor=south ]{$\mu_1\nu_1$};
		\filldraw [black] (0,-0.25) circle (0pt) node[anchor=south ]{$\mu_2\nu_2$};
		\filldraw [black] (-0.5,0.3) circle (0pt) node[anchor=north ]{\large $p$};
		\filldraw [black] (2,0) circle (0pt) node[anchor=north ]{ $=\langle h_{\mu_1\nu_1}(p) h_{\mu_2\nu_2}(-p)\rangle_{0}   $};
		\end{tikzpicture}
	\end{subfigure}
\nonumber\\
	\begin{subfigure}[b]{0.4\textwidth}
		\begin{tikzpicture}[scale=2.8,
		decoration={coil,amplitude=4.25,segment length=4.75},anchor=base, baseline]
		\draw[line width=1.2,black,  postaction={decorate, decoration={markings, mark=at position 1 with {\arrow[fill=black, scale=1.5]{Triangle}} } }] (-0.54,0)--(-0.541,0);
		\draw[line width=0.4,blue!7!black!80] (-.45,-0.025) -- (-1.3,-0.025);
		\draw[line width=0.4,blue!7!black!80] (-.532,0) -- (-1.3,0);
		\draw[line width=0.4,blue!7!black!80] (-.45,0.025) -- (-1.3,0.025);
		\draw[line width=0.4,blue!7!black!80] (-.47,-0.025) -- (.3,-0.025);
		\draw[line width=0.4,blue!7!black!80] (-.47,0) -- (.3,0);
		\draw[line width=0.4,blue!7!black!80] (-.47,0.025) -- (.3,0.025);
		\filldraw [black] (-1,0.05) circle (0pt) node[anchor=south ]{$ {\lambda_1} $};
		\filldraw [black] (-1,-0.25) circle (0pt) node[anchor=south ]{$ {\mu_1\nu_1} $};
		\filldraw [black] (0,0.05) circle (0pt) node[anchor=south ]{${\lambda_2}$};
		\filldraw [black] (0,-0.25) circle (0pt) node[anchor=south ]{${\mu_2\nu_2} $};
		\filldraw [black] (-0.5,0.3) circle (0pt) node[anchor=north ]{\large $p$};
		\filldraw [black] (2,0) circle (0pt) node[anchor=north ]{ $=\langle B^{\lambda_1}_{\mu_1\nu_1}(p) B^{\lambda_2}_{\mu_2\nu_2}(-p)\rangle_{0}   $};
		\end{tikzpicture}
	\end{subfigure}
	\nonumber\\
	\begin{subfigure}[b]{0.4\textwidth}
		\begin{tikzpicture}[scale=2.8,
		decoration={coil,amplitude=4.25,segment length=4.75},anchor=base, baseline]
		\draw[line width=1.2,black,  postaction={decorate, decoration={markings, mark=at position 1 with {\arrow[fill=black, scale=1.5]{Triangle}} } }] (-0.75,0)--(-0.76,0);
		\draw[line width=0.4,blue!7!black!80] (-.532,0) -- (-1.5,0);
		\draw[line width=0.4,blue!7!black!80] (-.67,0) -- (.1,0);
		\filldraw [black] (-1.2,0.05) circle (0pt) node[anchor=south ]{$ \mu_1 $};
		\filldraw [black] (-0.2,-0.25) circle (0pt) node[anchor=south ]{$\mu_2 $};
		\filldraw [black] (-0.7,0.3) circle (0pt) node[anchor=north ]{\large $p$};
		\filldraw [black] (2,0) circle (0pt) node[anchor=north ]{ $=\langle c^{\mu_1}(p)\bar{c}_{\mu_2}(p)\rangle_{0}   $};
		\end{tikzpicture}
	\end{subfigure}
	\caption{Free propagators} \label{fig:1}
\end{figure}
\begin{figure}
	\raggedright
	\begin{subfigure}[b]{0.4\textwidth}
		\begin{tikzpicture}[scale=2.8,
		decoration={coil,amplitude=4.25,segment length=4.75},anchor=base, baseline]
		\draw[line width=0.8,black,  postaction={decorate, decoration={markings, mark=at position 1 with {\arrow[fill=black, scale=1.5]{Triangle}} } }] (-0.441,0)--(-0.44,0);
		\draw[spring] (-.5,0) -- (-0.95,0);
		\draw[spring] (-0.45,0) -- (0,0);
		\draw[line width=0.8,black,  postaction={decorate, decoration={markings, mark=at position 1 with {\arrow[fill=black, scale=1.5]{Triangle}} } }] (0.31,0.31)--(0.3,0.3);
		\draw[spring] (.35,0.35) -- (0.6,0.6);
		\draw[spring] (0.305,0.305) -- (0,0);
		\draw[line width=0.8,black,  postaction={decorate, decoration={markings, mark=at position 1 with {\arrow[fill=black, scale=1.5]{Triangle}} } }] (0.31,-0.31)--(0.3,-0.3);
		\draw[spring] (.35,-0.35) -- (0.6,-0.6);
		\draw[spring] (0.305,-0.305) -- (0,0);
		\filldraw [black] (-0.5,-0.1) circle (0pt) node[anchor=north]{\large$p_1$};
		\filldraw [black] (-.5,0.05) circle (0pt) node[anchor=south ]{$\mu_1\nu_1 $};
		\filldraw [black] (.45,.33) circle (0pt) node[anchor=north ]{\large$p_2$};
		\filldraw [black] (.43,.4) circle (0pt) node[anchor=south east ]{$\mu_2\nu_2$};
		\filldraw [black] (.45,-.12) circle (0pt) node[anchor=north ]{\large$p_3$};
		\filldraw [black] (.43,-.65) circle (0pt) node[anchor=south east]{$\mu_3\nu_3$};
		\filldraw [black] (3.35,0) circle (0pt) node[anchor= east]{$=V_{(hhh)}^{\mu_1\nu_1,\mu_2\nu_2,\mu_3\nu_3}(p_1,p_2,p_3) $};
		\filldraw [black] (0,0) circle (1pt) node[anchor=south]{};
		\end{tikzpicture}
	\end{subfigure}
	\vspace{10mm}
	\nonumber\\
	\begin{subfigure}[b]{0.4\textwidth}
		\begin{tikzpicture}[scale=2.8,
		decoration={coil,amplitude=4.25,segment length=4.75},anchor=base, baseline]
		\draw[line width=0.8,black,  postaction={decorate, decoration={markings, mark=at position 1 with {\arrow[fill=black, scale=1.5]{Triangle}} } }] (-0.441,0)--(-0.44,0);
		\draw[spring] (-.5,0) -- (-0.95,0);
		\draw[spring] (-0.45,0) -- (0,0);
		\draw[line width=0.8,black,  postaction={decorate, decoration={markings, mark=at position 1 with {\arrow[fill=black, scale=1.5]{Triangle}} } }] (0.31,0.31)--(0.3,0.3);
		\draw[line width=0.4,blue!7!black!80] (.35,0.35) -- (0.6,0.6);
		\draw[line width=0.4,blue!7!black!80] (0.305,0.305) -- (0,0);
		\draw[line width=0.8,black,  postaction={decorate, decoration={markings, mark=at position 1 with {\arrow[fill=black, scale=1.5]{Triangle}} } }] (0.31,-0.31)--(0.3,-0.3);
		\draw[line width=0.4,blue!7!black!80] (.35,-0.35) -- (0.6,-0.6);
		\draw[line width=0.4,blue!7!black!80] (0.6,0.6) -- (0,0);
		\draw[line width=0.4,blue!7!black!80] (0.6,0.65) -- (0,0.05);
		\draw[line width=0.4,blue!7!black!80] (0.6,0.55) -- (0,-0.05);
		\draw[line width=0.4,blue!7!black!80] (0.305,-0.305) -- (0,0);
		\draw[line width=0.4,blue!7!black!80] (0.6,-0.6) -- (0,0);
		\draw[line width=0.4,blue!7!black!80] (0.6,-0.65) -- (0,-0.05);
		\draw[line width=0.4,blue!7!black!80] (0.6,-0.55) -- (0,0.05);
		\filldraw [black] (0,0) circle (1pt) node[anchor=south]{};
		\filldraw [black] (-0.5,-0.1) circle (0pt) node[anchor=north]{\large$p_3$};
		\filldraw [black] (-.5,0.05) circle (0pt) node[anchor=south ]{$\mu_3\nu_3 $};
		\filldraw [black] (.45,.33) circle (0pt) node[anchor=north ]{\large$p_1$};
		\filldraw [black] (.43,.46) circle (0pt) node[anchor=south east ]{$\mu_1\nu_1$};
		\filldraw [black] (.36,.28) circle (0pt) node[anchor=south east ]{$\l_1$};
		\filldraw [black] (.45,-.12) circle (0pt) node[anchor=north ]{\large$p_2$};
		\filldraw [black] (.43,-.65) circle (0pt) node[anchor=south east]{$\mu_2\nu_2$};
		\filldraw [black] (.41,-.84) circle (0pt) node[anchor=south east]{$\l_2$};
		\filldraw [black] (3.35,0) circle (0pt) node[anchor= east]{$=V^{~~~~~~\mu_1\nu_1,\mu_2\nu_2,\mu_3\nu_3}_{(BBh)~\l_1~~~\l_2}(p_1,p_2,p_3) $};
		\end{tikzpicture}
	\end{subfigure}
	\vspace{10mm}
	\nonumber\\
	\begin{subfigure}[b]{0.4\textwidth}
		\begin{tikzpicture}[scale=2.8,
		decoration={coil,amplitude=4.25,segment length=4.75},anchor=base, baseline]
		\draw[line width=0.8,black,  postaction={decorate, decoration={markings, mark=at position 1 with {\arrow[fill=black, scale=1.5]{Triangle}} } }] (-0.441,0)--(-0.44,0);
		\draw[line width=0.4,blue!7!black!80] (-.5,0.05) -- (-0.95,0.05);
		\draw[line width=0.4,blue!7!black!80] (-.5,0) -- (-0.95,0);
		\draw[line width=0.4,blue!7!black!80] (-.5,-0.05) -- (-0.95,-0.05);
		\draw[line width=0.4,blue!7!black!80](-0.5,0.05) -- (0,0.05);
		\draw[line width=0.4,blue!7!black!80](-0.45,0) -- (0,0);
		\draw[line width=0.4,blue!7!black!80](-0.5,-0.05) -- (0,-0.05);
		\draw[line width=0.8,black,  postaction={decorate, decoration={markings, mark=at position 1 with {\arrow[fill=black, scale=1.5]{Triangle}} } }] (0.31,0.31)--(0.3,0.3);
		\draw[spring] (.35,0.35) -- (0.6,0.6);
		\draw[spring] (0.305,0.305) -- (0,0);
		\draw[line width=0.8,black,  postaction={decorate, decoration={markings, mark=at position 1 with {\arrow[fill=black, scale=1.5]{Triangle}} } }] (0.31,-0.31)--(0.3,-0.3);
		\draw[spring] (.35,-0.35) -- (0.6,-0.6);
		\draw[spring] (0.305,-0.305) -- (0,0);
		\filldraw [black] (-0.5,-0.1) circle (0pt) node[anchor=north]{\large$p_1$};
		\filldraw [black] (-.5,0.2) circle (0pt) node[anchor=south ]{$\mu_1\nu_1 $};
		\filldraw [black] (-.5,0.05) circle (0pt) node[anchor=south ]{$\l_1 $};
		\filldraw [black] (.45,.33) circle (0pt) node[anchor=north ]{\large$p_2$};
		\filldraw [black] (.43,.4) circle (0pt) node[anchor=south east ]{$\mu_2\nu_2$};
		\filldraw [black] (.45,-.12) circle (0pt) node[anchor=north ]{\large$p_3$};
		\filldraw [black] (.43,-.65) circle (0pt) node[anchor=south east]{$\mu_3\nu_3$};
		\filldraw [black] (3.35,0) circle (0pt) node[anchor= east]{$=V^{~~~~~~\mu_1\nu_1,\mu_2\nu_2,\mu_3\nu_3}_{(Bhh)~~\l_1}(p_1,p_2,p_3) $};
		\filldraw [black] (0,0) circle (1pt) node[anchor=south]{};
		\end{tikzpicture}
	\end{subfigure}
	\vspace{10mm}
	\nonumber\\
	\begin{subfigure}[b]{0.4\textwidth}
		\begin{tikzpicture}[scale=2.8,
		decoration={coil,amplitude=4.25,segment length=4.75},anchor=base, baseline]
		\draw[line width=0.8,black,  postaction={decorate, decoration={markings, mark=at position 1 with {\arrow[fill=black, scale=1.5]{Triangle}} } }] (-0.441,0)--(-0.44,0);
		\draw[spring] (-.5,0) -- (-0.95,0);
		\draw[spring] (-0.45,0) -- (0,0);
		\draw[line width=0.8,black,  postaction={decorate, decoration={markings, mark=at position 1 with {\arrow[fill=black, scale=1.5]{Triangle}} } }] (0.31,0.31)--(0.3,0.3);
		\draw[line width=0.4,blue!7!black!80] (.35,0.35) -- (0.6,0.6);
		\draw[line width=0.4,blue!7!black!80] (0.305,0.305) -- (0,0);
		\draw[line width=0.8,black,  postaction={decorate, decoration={markings, mark=at position 1 with {\arrow[fill=black, scale=1.5]{Triangle}} } }] (0.3,-0.3)--(0.31,-0.31);
		\draw[line width=0.4,blue!7!black!80] (.35,-0.35) -- (0.6,-0.6);
		\draw[line width=0.4,blue!7!black!80] (0.6,0.6) -- (0,0);
		\draw[line width=0.4,blue!7!black!80] (0.305,-0.305) -- (0,0);
		\draw[line width=0.4,blue!7!black!80] (0.6,-0.6) -- (0,0);
		\filldraw [black] (0,0) circle (1pt) node[anchor=south]{};
		\filldraw [black] (-0.5,-0.1) circle (0pt) node[anchor=north]{\large$p_3$};
		\filldraw [black] (-.5,0.05) circle (0pt) node[anchor=south ]{$\mu_3\nu_3 $};
		\filldraw [black] (.45,.33) circle (0pt) node[anchor=north ]{$\m_2$};
		\filldraw [black] (.43,.46) circle (0pt) node[anchor=south east ]{\large$p_2$};
		\filldraw [black] (.45,-.12) circle (0pt) node[anchor=north ]{$\m_1$};
		\filldraw [black] (.43,-.65) circle (0pt) node[anchor=south east]{\large$p_1$};
		\filldraw [black] (3.35,0) circle (0pt) node[anchor= east]{$=V^{~~~~~~\mu_1,\mu_3\nu_3}_{(\bar{c}ch)~~\m_2}(p_1,p_2,p_3) $};
		\end{tikzpicture}
	\end{subfigure}
	\vspace{10mm}
	\nonumber\\
\begin{subfigure}[b]{0.4\textwidth}
	\begin{tikzpicture}[scale=2.8,
	decoration={coil,amplitude=4.25,segment length=4.75},anchor=base, baseline]
	\draw[line width=0.8,black,  postaction={decorate, decoration={markings, mark=at position 1 with {\arrow[fill=black, scale=1.5]{Triangle}} } }](0.5,0)-- (0.48,0);
	\draw[line width=0.4,blue!7!black!80] (0.5,0) -- (0.95,0);
	\draw[line width=0.4,blue!7!black!80](0,0) -- (0.48,0);
	\draw[line width=0.8,black,  postaction={decorate, decoration={markings, mark=at position 1 with {\arrow[fill=black, scale=1.5]{Triangle}} } }] (0.35,0.35)--(0.3,0.3);
	\draw[spring] (.35,0.35) -- (0.6,0.6);
	\draw[spring] (0.305,0.305) -- (0,0);
	\filldraw [black] (.45,.33) circle (0pt) node[anchor=north ]{\large$p_1$};
	\filldraw [black] (.43,.4) circle (0pt) node[anchor=south east ]{$\mu_1\nu_1$};
	\filldraw [black] (.45,-.12) circle (0pt) node[anchor=north ]{$\m_2$};
	\filldraw [black] (.65,.2) circle (0pt) node[anchor=north ]{\large$p_2$};
\filldraw [black] (-.33,0) circle (0pt) node[anchor=north ]{$\mu\nu$};
	\filldraw [black] (4,0) circle (0pt) node[anchor= east]{$=V^{~~~~~~\mu\nu,\mu_1\nu_1}_{(RHO)~~~~~~~~\m_2}(p_1,p_2) $};
	\draw (0,0) circle (2pt);
	\draw (0,0) node[cross=5pt,rotate=90]{};
	\end{tikzpicture}
\end{subfigure}
	\caption{Vertices}\label{fig:2}
\end{figure}

\newpage
\section{The Slavnov-Taylor identity for the 1PI functional}\label{B}
The simplest interpretation of the Slavnov-Taylor \cite{Slavnov} identities is that they just represent the invariance of the 1PI effective action with respect to the nilpotent BRST transformations. As such, they are a powerful check of the gauge invariance of the final results.
\par
 Let us see this in some detail.
\par
Let $\rho_{\mu\nu}(x)$, $\rho^{\mu\nu}_\lambda(x)$ and $\sigma_\mu(x)$ be external fields which couple to the BRST transformations of $h^{\mu\nu}(x)$, $B^{\lambda}_{\mu\nu}(x)$ and $c^{\mu}(x)$, respectively. One defines the BRST variation of those external fields as follows
\begin{equation*}
s\rho_{\mu\nu}(x)=0,\quad s\rho^{\mu\nu}_\lambda(x)=0, s\sigma_\mu(x)=0.
\end{equation*}
Let $\Gamma_0[h^{\mu\nu},B_{\mu\nu}^{\lambda},b_\mu,c^\mu,\bar{c}_\mu;\rho_{\mu\nu},\rho^{\mu\nu}_\lambda]$ be given by
\begin{equation}
\Gamma_0[h^{\mu\nu},B_{\mu\nu}^{\lambda},b_\mu,c^\mu,\bar{c}_\mu;\rho_{\mu\nu},\rho^{\mu\nu}_\lambda]=
S[h^{\mu\nu},B_{\mu\nu}^{\lambda}]+sX+i\int d^n x\,(\rho_{\mu\nu}sh^{\mu\nu}+\rho^{\mu\nu}_\lambda sB^\lambda_{\mu\nu}+\sigma_\mu sc^\mu),
\label{1PI0}
\end{equation}
where $S[h^{\mu\nu},B_{\mu\nu}^{\lambda}]$ is the classical action of the theory without the gauge-fixing term. Clearly,
\begin{equation}
s\Gamma_0[h^{\mu\nu},B_{\mu\nu}^{\lambda},b_\mu,c^\mu,\bar{c}_\mu;\rho_{\mu\nu},\rho^{\mu\nu}_\lambda]=0.
\label{sgamma0}
\end{equation}

Then, the generating functional of the  Green functions of the theory, including the insertions of the BRST variations which are given by composite operators, reads
\begin{equation*}
\begin{array}{l}
{Z[j_{\mu\nu},j^{\mu\nu}_{\lambda},j^\mu,\bar{\omega}_\mu,\omega^\mu;\rho_{\mu\nu},\rho^{\mu\nu}_\lambda]=}\\[8pt]
{{\cal N}\int{\cal D}h^{\mu\nu}{\cal D}B^\lambda_{\mu\nu}{\cal D}b^{\mu}{\cal D}c^{\mu}{\cal D}\bar{c}_{\mu}\,
e^{i\Gamma_0+i\int d^n x\,\left[j_{\mu\nu}h^{\mu\nu}+j_{\lambda}^{\mu\nu}B^{\lambda}_{\mu\nu}+j^\mu b_\mu+\bar{\omega}_\mu c^{\mu}+\bar{c}_\mu \omega^\mu\right]}.
}
\end{array}
\end{equation*}

Let us denote by
\begin{equation*}
\Gamma\equiv\Gamma[h^{\mu\nu},B_{\mu\nu}^{\lambda},b_\mu,c^\mu,\bar{c}_\mu;\rho_{\mu\nu},\rho^{\mu\nu}_\lambda]
\end{equation*}
the 1PI functional deduced from
$Z[j_{\mu\nu},j^{\mu\nu}_{\lambda},j^\mu,\bar{\omega}_\mu,\omega^\mu;\rho_{\mu\nu},\rho^{\mu\nu}_\lambda]$ above.
By doing standard manipulations -see \cite{Becchi:1996yh}, one can show that the fact that (\ref{sgamma0}) holds leads to the conclusion that the 1PI functional $\Gamma$ must satisfy the so-called Slavnov-Taylor identity:
\begin{equation}
\int d^n x\,\left[
\cfrac{\d\Gamma}{\d \rho_{\m\n}}\cfrac{\d\Gamma}{\d h^{\m\n}}+\cfrac{\d\Gamma}{\d \rho^{\m\n}_\lambda}\cfrac{\d\Gamma}{\d B^\lambda_{\m\n}}
+\cfrac{\d\Gamma}{\d \sigma_\m}\cfrac{\d\Gamma}{\d c^\m}+b_\m\cfrac{\d\Gamma}{\d \bar{c}_\m}\right]=0.
\label{ST1PI}
\end{equation}
This is the equation which governs the BRST symmetry at the quantum level.

For the choice of gauge-fixing done in this paper --see section 2-- the Slavnov-Taylor equation can be simplified a bit. Indeed, it is not difficult to show for our gauge-fixing choice, we have
\begin{equation*}
\cfrac{\d\Gamma}{\d b_\m(x)}=b^\mu(x)-\partial_{\nu}h^{\nu\mu}(x).
\end{equation*}
Then, it is sensible to introduce
\begin{equation}
\tilde{\Gamma}=\Gamma-\int d^n x \left(\cfrac{1}{2}\,b_\mu b^\mu-b_\mu\partial_\nu h^{\nu\mu}\right),
\label{t1PI}
\end{equation}
so that
\begin{equation*}
\cfrac{\d\tilde{\Gamma}}{\d b_\m(x)}=0.
\end{equation*}
By substituting (\ref{t1PI}) in (\ref{ST1PI}), one concludes that the Slavnov-Taylor equation above is equivalent to the following set equations
\begin{equation}
\begin{array}{l}
{\int d^n x\,\left[
\cfrac{\d\tilde{\Gamma}}{\d \rho_{\m\n}}\cfrac{\d\tilde{\Gamma}}{\d h^{\m\n}}+\cfrac{\d\tilde{\Gamma}}{\d \rho^{\m\n}_\lambda}\cfrac{\d\tilde{\Gamma}}{\d B^\lambda_{\m\n}}
+\cfrac{\d\tilde{\Gamma}}{\d \sigma_\m}\cfrac{\d\tilde{\Gamma}}{\d c^\m}\right]=0,}\\[12pt]
{\partial_\mu\cfrac{\d\tilde{\Gamma}}{\d \rho_{\m\n}}-\cfrac{\d\tilde{\Gamma}}{\d \bar{c}_\n}=0.}
\end{array}
\label{the2eq}
\end{equation}
The second equation is equivalent to the statement that $\rho_{\mu\nu}$ and $\bar{c}_\m$ always occur in $\tilde{\Gamma}$ in the combination
\begin{equation*}
\rho_{\mu\nu}-\cfrac{1}{2}(\partial_\mu\bar{c}_\nu+\partial_\nu\bar{c}_\mu),
\end{equation*}
which we shall call $\tilde{\rho}_{\m\n}$. In terms of this field, the first equation in (\ref{the2eq}) reads
\begin{equation*}
\int d^n x\,\left[
\cfrac{\d\tilde{\Gamma}}{\d \tilde{\rho}_{\m\n}}\cfrac{\d\tilde{\Gamma}}{\d h^{\m\n}}+\cfrac{\d\tilde{\Gamma}}{\d \rho^{\m\n}_\lambda}\cfrac{\d\tilde{\Gamma}}{\d B^\lambda_{\m\n}}
+\cfrac{\d\tilde{\Gamma}}{\d \sigma_\m}\cfrac{\d\tilde{\Gamma}}{\d c^\m}\right]=0.
\end{equation*}

By expanding $\tilde{\Gamma}$ in the number of loops, one reaches the conclusion that the one-loop contribution, say $\tilde{\Gamma}_1$, to that 1PI functional must satisfy the so-called linearized Slavnov-Taylor identity:
\begin{equation}
\int d^n x\,\left[
\cfrac{\d\tilde{\Gamma}_0}{\d \tilde{\rho}_{\m\n}}\cfrac{\d\tilde{\Gamma}_1}{\d h^{\m\n}}+
\cfrac{\d\tilde{\Gamma}_0}{\d h^{\m\n}}\cfrac{\d\tilde{\Gamma}_1}{\d \tilde{\rho}_{\m\n}}
+\cfrac{\d\tilde{\Gamma}_0}{\d \rho^{\m\n}_\lambda}\cfrac{\d\tilde{\Gamma}_1}{\d B^\lambda_{\m\n}}+
\cfrac{\d\tilde{\Gamma}_0}{\d B^\lambda_{\m\n}}\cfrac{\d\tilde{\Gamma}_1}{\d \rho^{\m\n}_\lambda}
+\cfrac{\d\tilde{\Gamma}_0}{\d \sigma_\m}\cfrac{\d\tilde{\Gamma}_1}{\d c^\m}+
\cfrac{\d\tilde{\Gamma}_0}{\d c^\m}\cfrac{\d\tilde{\Gamma}_1}{\d \sigma_\m}
\right]=0,
\label{linearizedST}
\end{equation}
where
\begin{equation*}
\tilde{\Gamma}_0=\Gamma_0-\int d^n x \left[\cfrac{1}{2}\,b_\mu b^\mu-b_\mu\partial_\nu h^{\nu\mu}\right].
\label{tildegamma}
\end{equation*}
$\Gamma_0$ being defined (\ref{1PI0}).

Now, by computing the following second derivative
\begin{equation*}
\cfrac{\d^2\phantom{c^{\rho}(x)}}{\d c^{\rho}(x)\d h^{\m\n}(y)}
\end{equation*}
of (\ref{linearizedST}) and then setting all the fields to zero, one obtains, upon Fourier transform, the identity in (\ref{SThh}). The reader should recall that $s B^{\lambda}_{\mu\nu}$, as defined in (\ref{Bvariation}), has no contribution involving only the field  $c^{\mu}$.


\end{document}